\documentclass[12pt,preprint]{aastex}

\usepackage{graphicx,epsfig,subfigure,rotating}
\usepackage{amssymb,amsmath}

\begin{document}


\title{The impact of point source subtraction residuals on 21~cm Epoch of Reionization estimation}


\author{Cathryn M. Trott, Randall B. Wayth and Steven J. Tingay}
\affil{International Centre for Radio Astronomy Research, Curtin University, Bentley WA, Australia}
\affil{ARC Centre of Excellence for All-Sky Astrophysics (CAASTRO)}
\email{cathryn.trott@curtin.edu.au}




\begin{abstract}
Precise subtraction of foreground sources is crucial for detecting and estimating 21~cm HI signals from the Epoch of Reionization (EoR). We quantify how imperfect point source subtraction due to limitations of the measurement dataset yields structured residual signal in the dataset. We use the Cramer-Rao lower bound, as a metric for quantifying the precision with which a parameter may be measured, to estimate the residual signal in a visibility dataset due to imperfect point source subtraction. We then propagate these residuals into two metrics of interest for 21~cm EoR experiments -- the angular power spectrum and two-dimensional power spectrum -- using a combination of full analytic covariant derivation, analytic variant derivation, and covariant Monte Carlo simulations. This methodology differs from previous work in two ways: (1) it uses information theory to set the point source position error, rather than assuming a global root-mean-square error, and (2) it describes a method for propagating the errors analytically, thereby obtaining the full correlation structure of the power spectra. The methods are applied to two upcoming low-frequency instruments that are proposing to perform statistical EoR experiments: the Murchison Widefield Array (MWA) and the Precision Array for Probing the Epoch of Reionization (PAPER). In addition to the actual antenna configurations, we apply the methods to minimally-redundant and maximally-redundant configurations. We find that for peeling sources above 1~Jy, the amplitude of the residual signal, and its variance, will be smaller than the contribution from thermal noise for the observing parameters proposed for upcoming EoR experiments, and that optimal subtraction of bright point sources will not be a limiting factor for EoR parameter estimation. We then use the formalism to provide an \textit{ab initio} analytic derivation motivating the ``wedge" feature in the two-dimensional power spectrum, complementing previous discussion in the literature.
\end{abstract}




\section{Introduction}
An understanding of the systematic observational biases in a dataset is crucial for forming an unbiased estimate of the Epoch of Reionization (EoR) signal, as measured by low-frequency interferometric radio telescopes, because the signal is expected to be weak (orders of magnitude below contaminating signals and the thermal noise), and the scientific use of the result relies on the detailed Fourier structure of the signal. Foreground contaminants to the EoR 21~cm signal include bright point sources, confusing (unresolved) point sources, and diffuse galactic synchrotron emission. Studies have been undertaken to explore methods for removing unresolved sources and diffuse structure, and the effects of these techniques on the signal, but it is typically assumed that the bright source subtraction has been performed perfectly \citep[][and references therein]{bowman09,liu09,liu11}. These bright sources are often used to calibrate the instrument \citep[e.g., Murchison Widefield Array -- MWA{\footnote[1]{http://www.mwatelescope.org}}, Precision Array for Probing the Epoch of Reionoization -- PAPER{\footnote[2]{http://eor.berkeley.edu}}, Low Frequency Array -- LOFAR{\footnote[3]{http://www.lofar.org}}, Long Wavelength Array -- LWA{\footnote[4]{http://lwa.unm.edu}},][]{lonsdale09,tingay12,parsons10,stappers11,ellingson09}. Imperfect source subtraction may lead to systematic errors in the dataset used to perform EoR measurement. It is this effect we study in this paper.

There have been previous attempts to quantify the uncertainties introduced into an image dataset, and consequently a measure of the EoR power spectrum \citep{datta09,datta10}, by position errors in the bright point source subtraction. These studies have assumed position errors with a global root-mean-square magnitude, and propagated them to the final measurement. They have not studied the expected position errors introduced by imprecise measurement of the point sources (i.e., uncertainties in parameter estimation due to imperfect data). An analysis that derives the expected parameter uncertainties, and then propagates them analytically into the visibility and image datasets, will provide the most reliable estimate of the impact of bright source contamination. In this paper we present a method for determining the impact of bright source subtraction from first principles. Our method provides a straight-forward analysis, based on the Cramer-Rao lower bounds (CRBs), of the theoretical level to which the parameters of a source, and therefore our ability to subtract it, can be determined, and propagates these uncertainties into the visibility plane and EoR power spectra.

Bright point sources affect interferometric datasets substantially. In addition to saturation of signal within a synthesized beam-width of the actual source, the incomplete $uv$ coverage of distributed synthesis instruments leads to strong sidelobe contamination throughout the dataset. For measurement of the weak ($\sim$20--50~mK) and structured EoR signal, bright point sources ($>$1~Jy) need to be accurately and precisely removed from the dataset before analysis can proceed. Precise estimates of the source parameters are achievable due to the strong signal from these sources (high signal-to-noise ratio). Conversely, inaccurate subtraction of the brightest sources may lead to large amplitude residuals. The balance between these factors will determine the utility of these datasets for EoR science.

At low frequencies ($\lesssim$200~MHz), the ionosphere produces a measurable perturbative effect on the wavefront shape of a propagating signal, even for short baseline arrays specifically targetting the EoR (e.g., MWA, PAPER). The ionosphere acts as a phase screen. The first order gradient term yields a shift in source position, while higher-order terms cause lensing. Spatially compact arrays see very little curvature, making a gradient phase screen a reasonable approximation under normal ionospheric conditions. \citet{cohen09} studied the differential refraction (position change relative to other sources in the field) for Very Large Array (VLA) fields at 74~MHz. They found an empirical power law fit to the differential refraction as a function of source sky separation of $\sim$0.5 during nighttime observations ($\sim$0.7 during the day), under normal ionospheric conditions (refraction regime), yielding a source shift of $\sim$100 arcseconds at a separation of 25 degrees. In addition to this, the whole field suffers an overall shift of $\sim$1--10 arcminutes. At 150~MHz, one would expect the refraction to be half the size, owing to the frequency dependence of phase rotation.

Wide-field low frequency instruments, such as the MWA, PAPER and LWA, aim to use known bright point sources as calibrators to model the instantaneous effect of the ionosphere on the wavefront. In addition to these calibration sources, lower flux sources will need to be removed from the dataset for EoR signal estimation. In this work we consider an observational scheme where the calibration is performed in real-time, necessitating a measurement of the instrument and sky response on short timescales, and applying these measurements in real-time to the measured data (visibilities). We refer to peeled sources generically as ``calibrators'', although some sources may be peeled as contaminating foregrounds, but not used for instrument calibration.

\citet{mitchell08} list the steps in the MWA Calibration Measurement Loop (CML) within the Real-time System (RTS). The CML is designed to track the real-time wavefront distortions due to the ionosphere, by performing instrumental calibration on short timescales ($\sim$8--10 seconds). The calibration is performed on bright calibrators, which are then subtracted (``peeled") from the visibility dataset before further data processing. The potential for ionospheric changes over the calibration timescale effectively corresponds to a re-measurement of the parameters of each point source each 8--10 seconds. Accurate and precise subtraction of these calibrators is therefore limited by the amount of information available about the parameters of a source (brightness and sky position) in that period, yielding a lower limit to the precision with which estimation (and therefore subtraction) can occur. For an unbiased estimator, this fundamental limit is set by the Cramer-Rao bound (CRB) on parameter estimation \citep{kay93}. We will use the CRB to quantify the maximal precision with which the parameters (sky position and flux density) of a source may be estimated, and use these uncertainties to model the residual point source signal in the visibility dataset. We then describe a formalism for propagating these errors to power spectra, using a fully-covariant analytic derivation. We use this formalism, along with Monte-Carlo simulations to estimate the residual power and uncertainty in the angular power spectrum and two-dimensional power spectrum due to point source subtraction. Note that this approach differs from that of \citet{datta10} in two ways: (1) it uses information theory to set the point source position error, rather than assuming a global error; (2) it propagates the full covariance matrix to the power spectra estimates, yielding both the uncertainties on the power values, and the level of correlation between $k$-modes. We work from first principles in our approach.

\section{Methods}

\subsection{Estimation performance: Cramer-Rao lower bound}\label{crb_section}
To determine the theoretical optimal estimation performance with a given dataset, we can calculate the Cramer-Rao lower bound (CRB) on the precision of parameter estimates. The CRB calculates the precision with which a minimum-variance unbiased estimator could estimate a parameter value, \textit{using the information content of the dataset}. It is computed as the square-root of the corresponding diagonal element of the inverse of the Fisher information matrix (FIM). The ($ij$)th entry of the FIM for a vector ${\boldsymbol{\theta}}$ of unknown parameters is given by:
\begin{equation}
[\boldsymbol{I(\theta)}]_{ij} = -\textless\left[\frac{\partial^2{\log{L({\bf{x}};{\boldsymbol{\theta}})}}}{\partial{\theta_i}\partial{\theta_j}} \right]\textgreater,
\end{equation}
where $L$ denotes the likelihood function describing the likelihood of measuring a dataset, ${\bf x}$ for a given parameter set, $\boldsymbol{\theta}$. For a general covariance matrix, $\boldsymbol{C}$, $N$ independent samples and complex data modelled with signal $\tilde{\boldsymbol{s}}$, the general expression for the Fisher Information Matrix is:
\begin{eqnarray}
[\boldsymbol{I(\theta)}]_{ij} = {\rm{tr}}\left[ \boldsymbol{C}^{-1}(\boldsymbol{\theta})\frac{\partial\boldsymbol{C}(\boldsymbol{\theta})}{\partial{\theta_i}}\boldsymbol{C}^{-1}(\boldsymbol{\theta})\frac{\partial\boldsymbol{C}(\boldsymbol{\theta})}{\partial{\theta_j}} \right] \\\nonumber
+ 2{\rm{Re}}\left[\frac{\partial{\tilde{\boldsymbol{s}}^\dagger({\boldsymbol{\theta}})}}{\partial{\theta_i}}{\boldsymbol{C}^{-1}}(\boldsymbol{\theta})\frac{\partial{\tilde{\boldsymbol{s}}({\boldsymbol{\theta}})}}{\partial{\theta_j}}\right],
\end{eqnarray}
simplifying to,
\begin{equation}
[\boldsymbol{I(\theta)}]_{ij} = 2{\rm{Re}}\left[\frac{1}{\sigma^2}\displaystyle\sum_{n=1}^N\frac{\partial{\tilde{s}^\dagger[n;{\boldsymbol{\theta}}]}}{\partial{\theta_i}}\frac{\partial{\tilde{s}[n;{\boldsymbol{\theta}}]}}{\partial{\theta_j}}\right]
\label{CRB}
\end{equation}
for white Gaussian noise with variance, $\sigma^2$ \citep{kay93}.

The CRB is a useful metric because it places a fundamental lower limit on the measurement precision of any parameter. In this work it will be used to gain an understanding of the fundamental limits of point source subtraction, and how these impact EoR estimation. We have previously used it to quantify the precision with which parameters of a point source may be estimated with an interferometric visibility dataset \citep{trott11}. Note that the results in this work assume an unbiased and efficient estimator exists, in which case the CRB yields the ideal precision. In practise, such an estimator may not exist, and the bound will be unattainable. Precision better than the CRB can be achieved with a biased estimator.

\subsection{Calibration field: point source population}
To model a realistic distribution of calibrators, we use the differential source counts of \citet{hales88}, obtained from the 6C survey at 151~MHz:
\begin{equation}
\frac{dN}{dS} = 3600S^{-2.5}_{\rm Jy} {\rm Jy}^{-1} {\rm str}^{-1},
\label{dNdS}
\end{equation}
A minimum flux density of 1~Jy will be assumed. We generate a catalogue of flat-spectrum calibrators according to equation \ref{dNdS}, with sky positions assigned randomly across the field, yielding $\sim$400 calibration sources in a 30$\times$30 degree$^2$ field ($\sim$1600 in a 60$\times$60 field), with a maximum flux density of 40--60~Jy.

\subsection{Instrument design}


We focus our attention on the MWA instrument, currently under construction in Western Australia, with science commissioning expected to commence in the second half of 2012. The MWA is composed of 128 electronically-steerable tiles, each of which comprises a ground plane with sixteen fixed dipole antennas \citep{lonsdale09,mitchell08,tingay12}. The maximum baseline will be $\sim$3 kilometres, with an instantaneous observing bandwidth of 30.72~MHz over the 80--300~MHz range. The MWA will boast a wide field-of-view ($\sim$30 degrees at 150~MHz) and high sensitivity, making it an ideal instrument for detection and measurement of the 21~cm Epoch of Reionization signal at $z\sim$~6--11. Table \ref{instrument_design} lists the fiducial MWA parameters used in this work ($\sigma_{\rm RMS}$ is the theoretical noise level per visibility, using the whole instrumental bandwidth).
\begin{table}
\begin{center}
\begin{tabular}{|c|c|c|}
\hline Parameter & Value (MWA) & Value (PAPER) \\ 
\hline\hline N$_{\rm ant}$ & 128 & 64 \\ 
\hline $\nu_0$ & 150 MHz & 150 MHz \\
\hline Bandwidth & 30.72 MHz & 70.0 MHz \\
\hline $\Delta\nu_{\rm 1D}$ & 125 kHz & 125 kHz \\
\hline $\Delta\nu_{\rm 2D}$ & 8 MHz & 8 MHz \\
\hline Field-of-view & 30$^o$ & 60$^o$ \\ 
\hline Calibrators ($>$1 Jy) & 392 & 1579\\
\hline T$_{\rm sys}$ & 440K & 440K \\ 
\hline $\Delta{t}$ & 8s & 8s \\ 
\hline $\sigma_{\rm RMS}$ & 28.1~mK & 18.6~mK \\
\hline $t_{\rm tot}$ & 300h & 720h \\ 
\hline $d_{\rm max}$ & 3000m & 260m \\ 
\hline
\end{tabular}
\caption{Parameters used for the primary instrument design of the MWA and PAPER 64-dipole instruments. For all experiments we use the full instantaneous bandwidth of the instrument to estimate source parameter precision.}\label{instrument_design}
\end{center}
\end{table}
We use the MWA array configuration of \citet{beardsley12} for the antenna layout.

\begin{figure}
\begin{center}
\subfigure[MWA -- uniform, antenna locations.]{\includegraphics[scale=0.32]{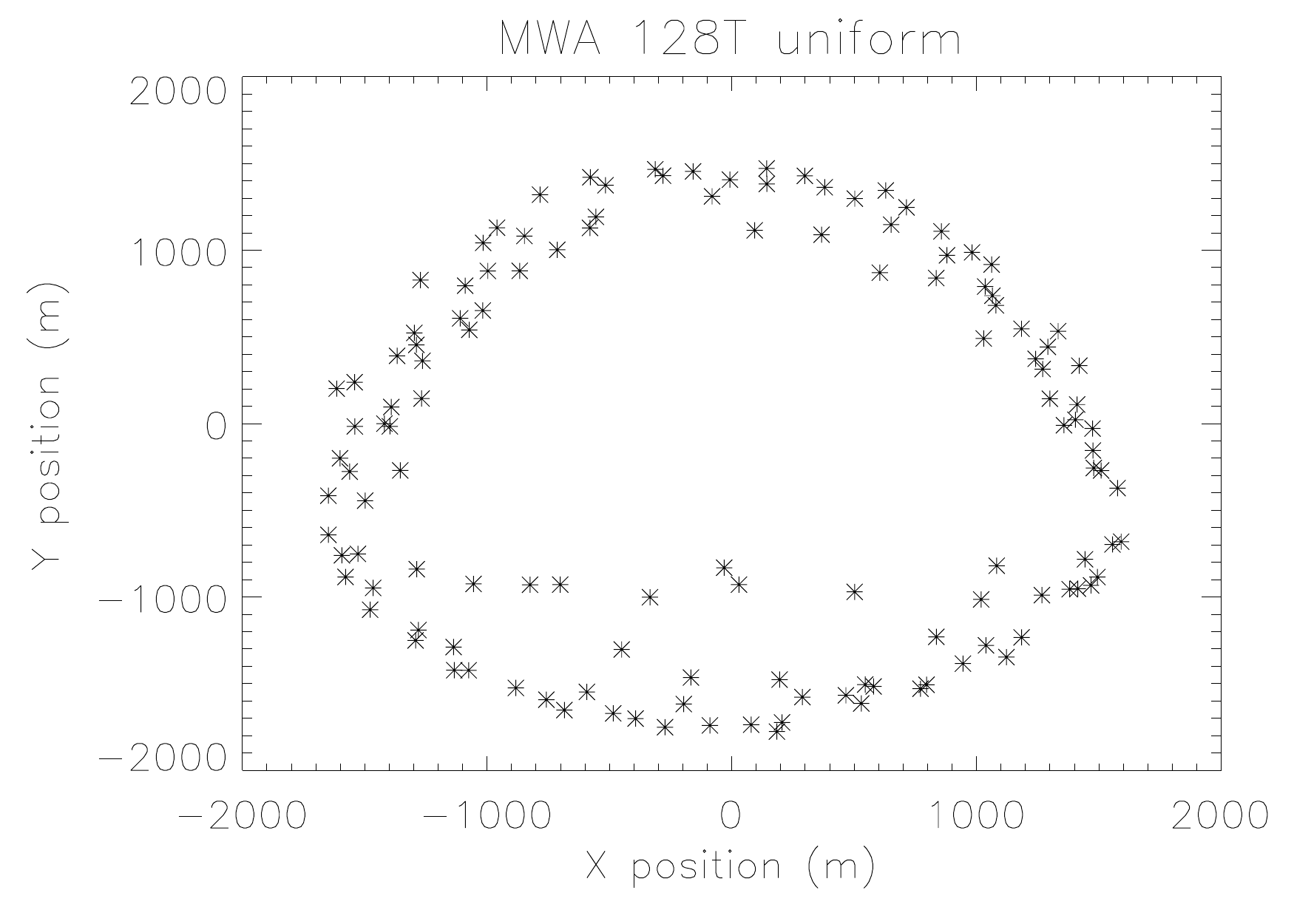}}
\subfigure[MWA -- uniform, $uv$ coverage.]{\includegraphics[scale=0.32]{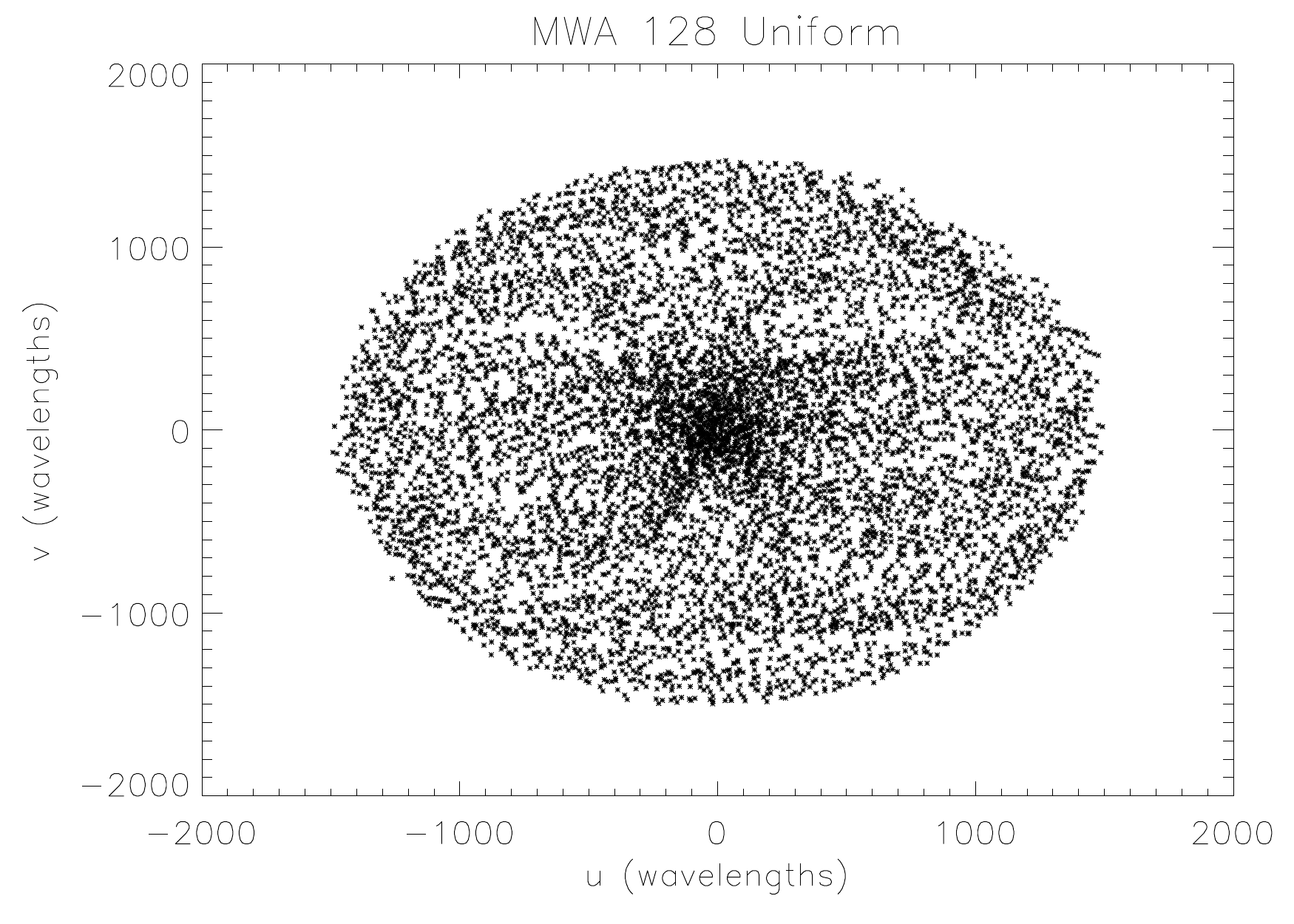}}\\
\subfigure[MWA, antenna locations.]{\includegraphics[scale=0.32]{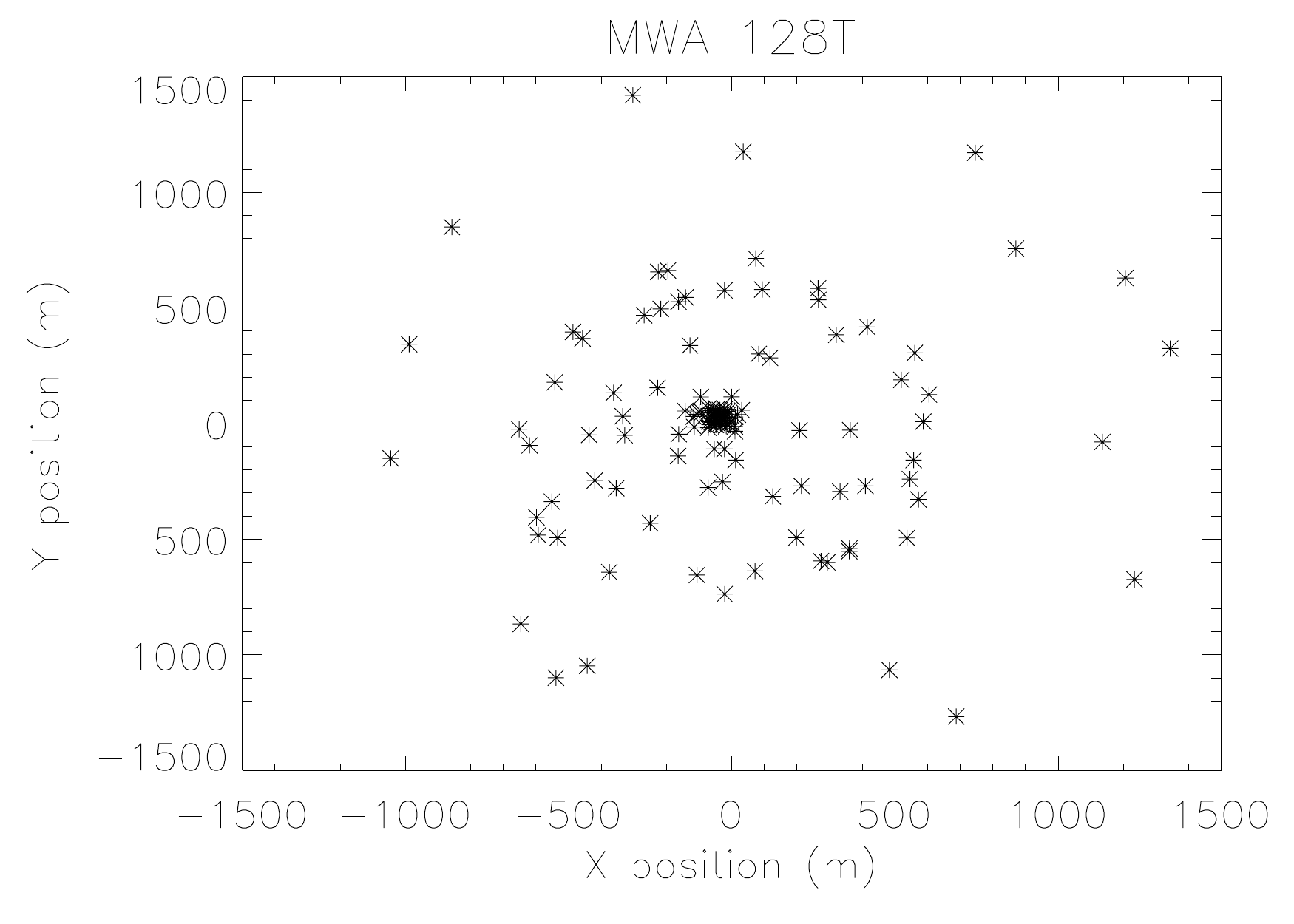}}
\subfigure[MWA, $uv$ coverage.]{\includegraphics[scale=0.32]{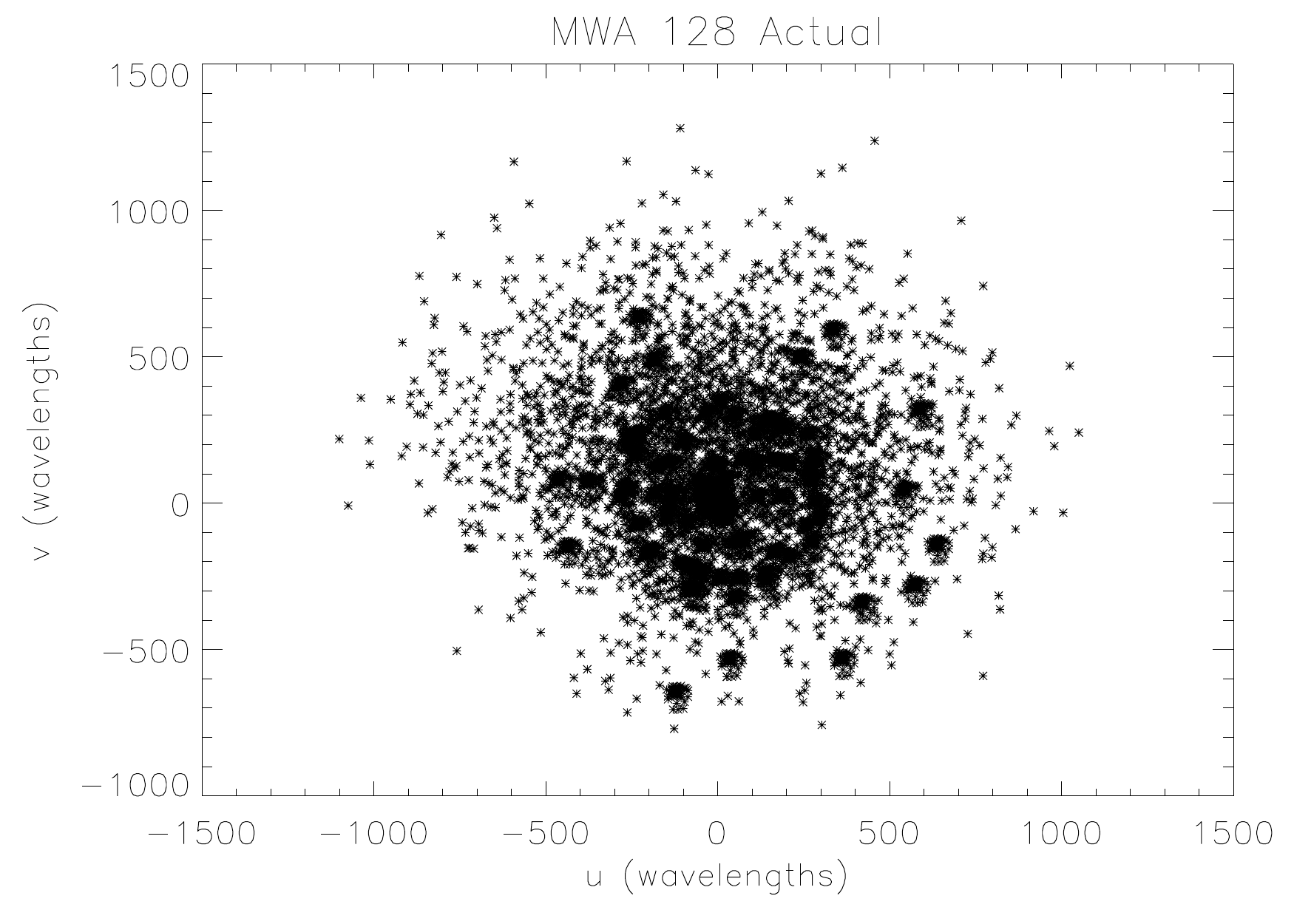}}\\
\subfigure[PAPER minimum-redundancy, antenna locations.]{\includegraphics[scale=0.32]{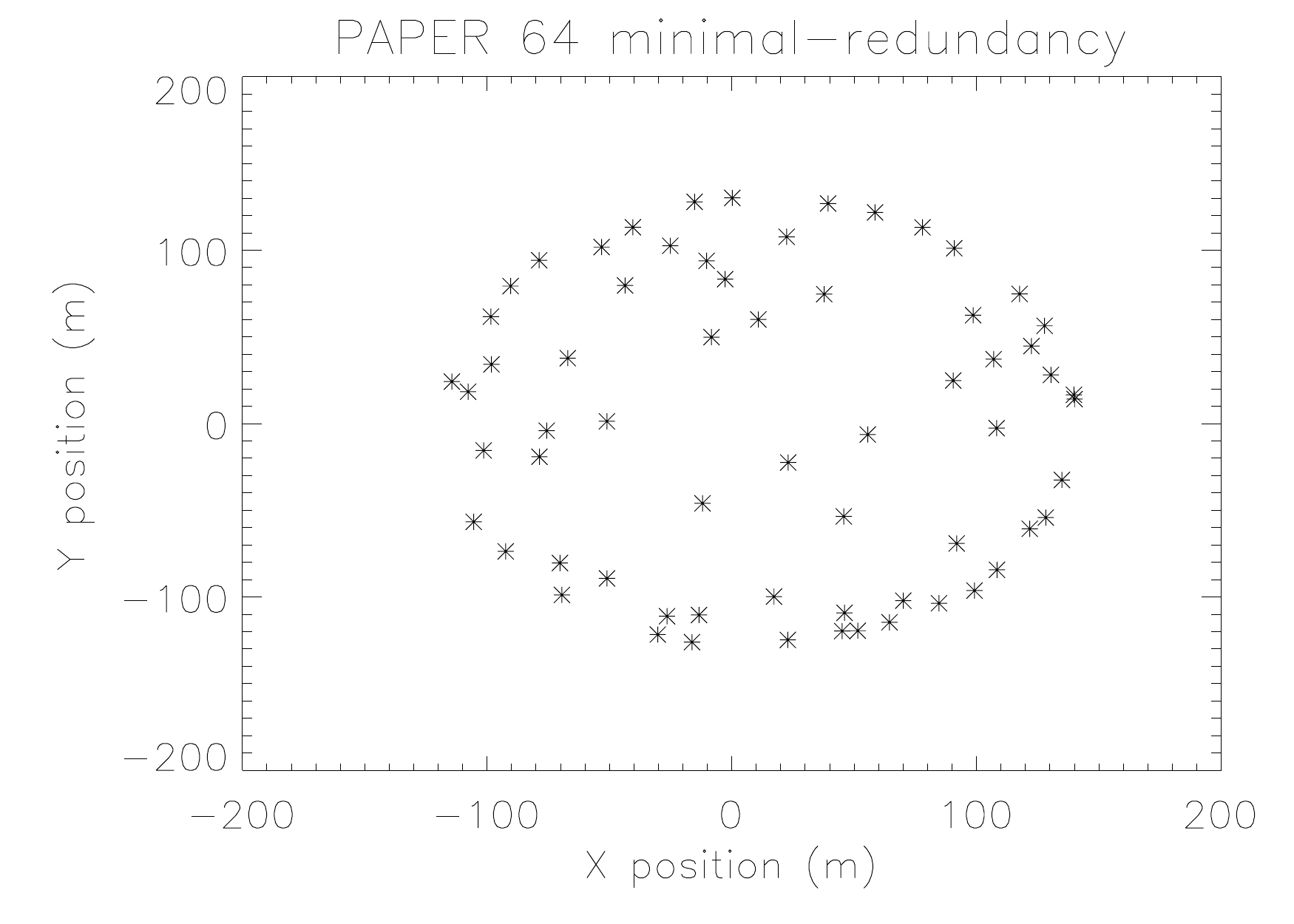}}
\subfigure[PAPER minimum-redundancy, $uv$ coverage.]{\includegraphics[scale=0.32]{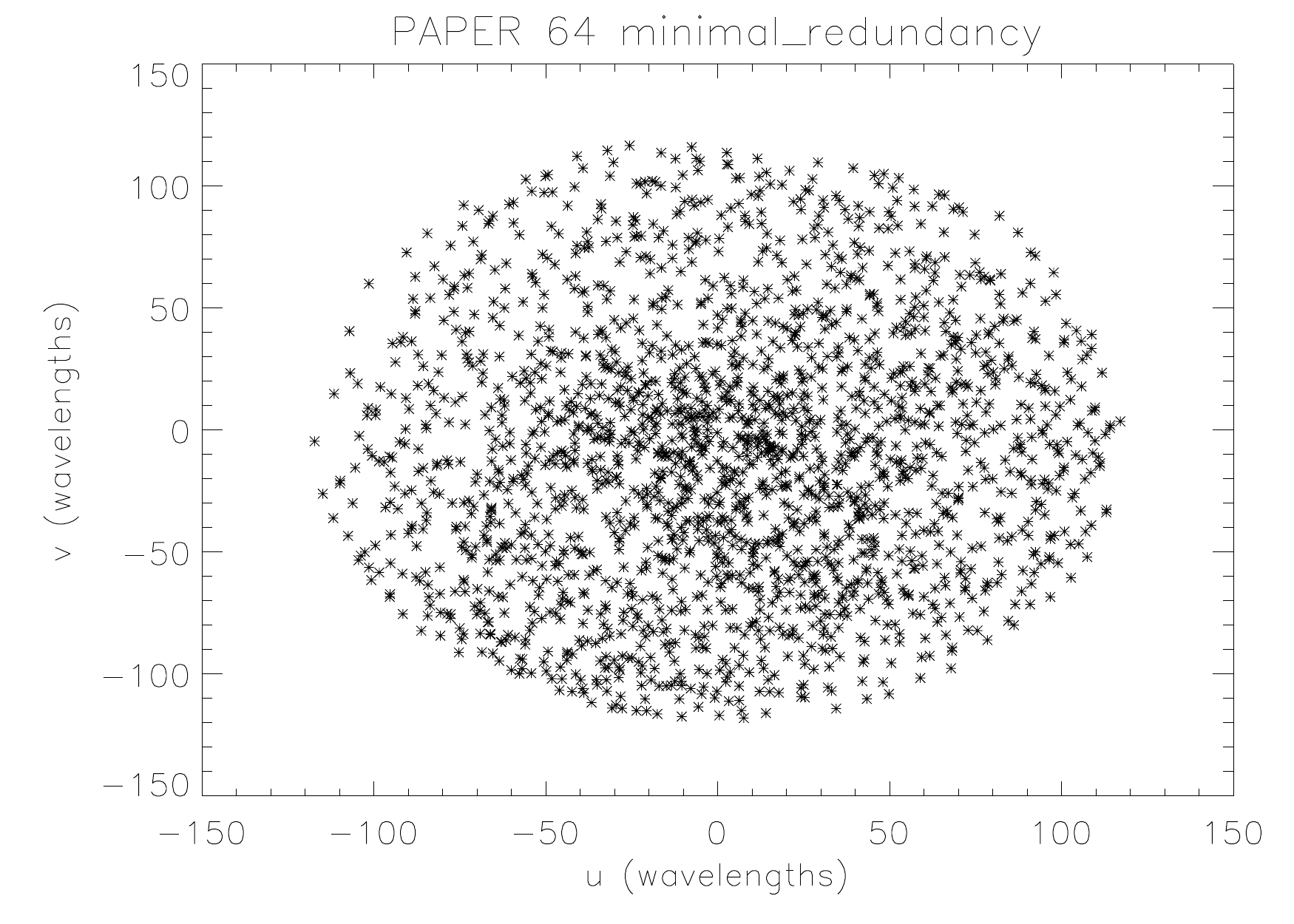}}\\
\subfigure[PAPER maximum-redundancy, antenna locations.]{\includegraphics[scale=0.32]{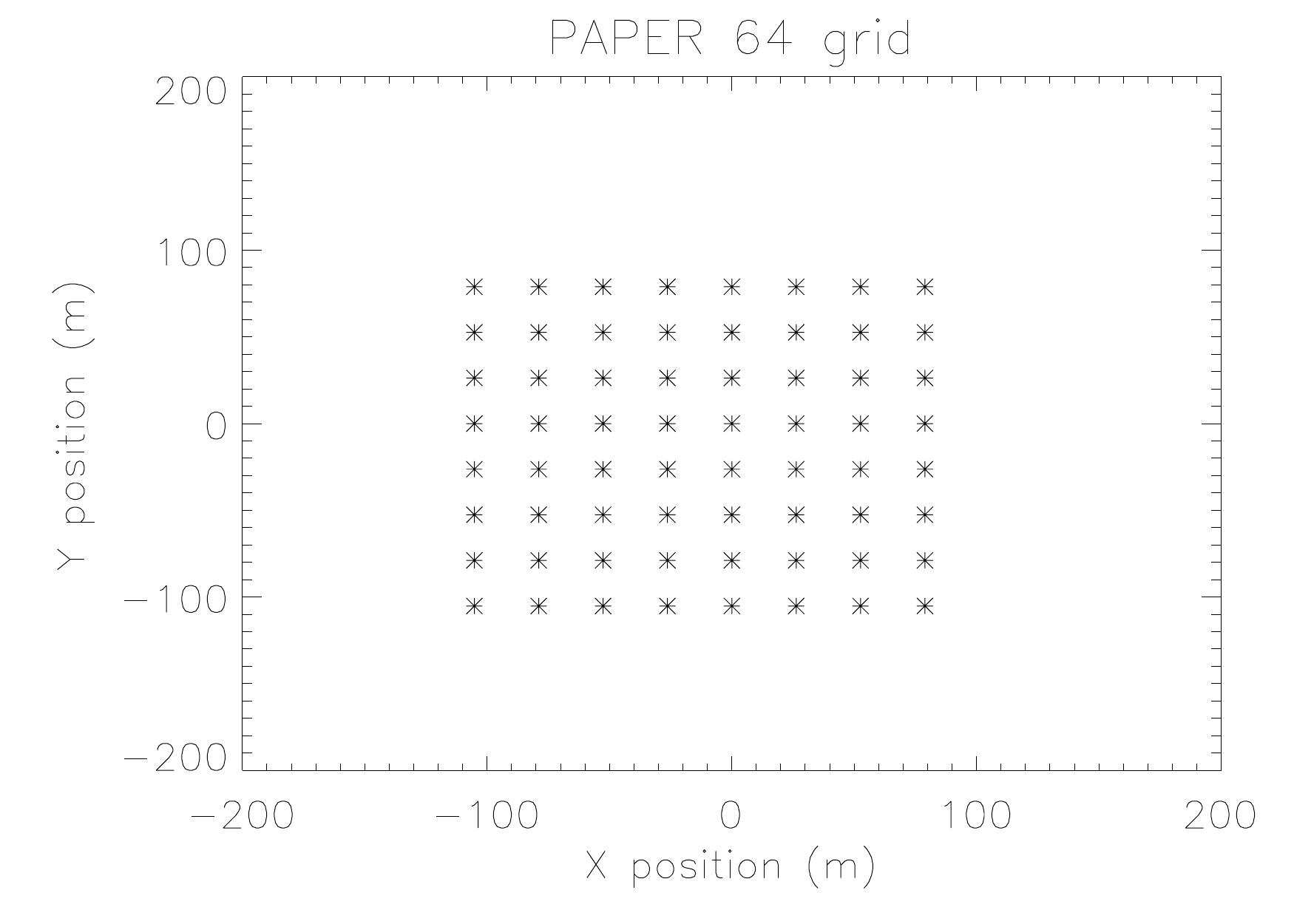}}
\subfigure[PAPER maximum-redundancy, $uv$ coverage.]{\includegraphics[scale=0.32]{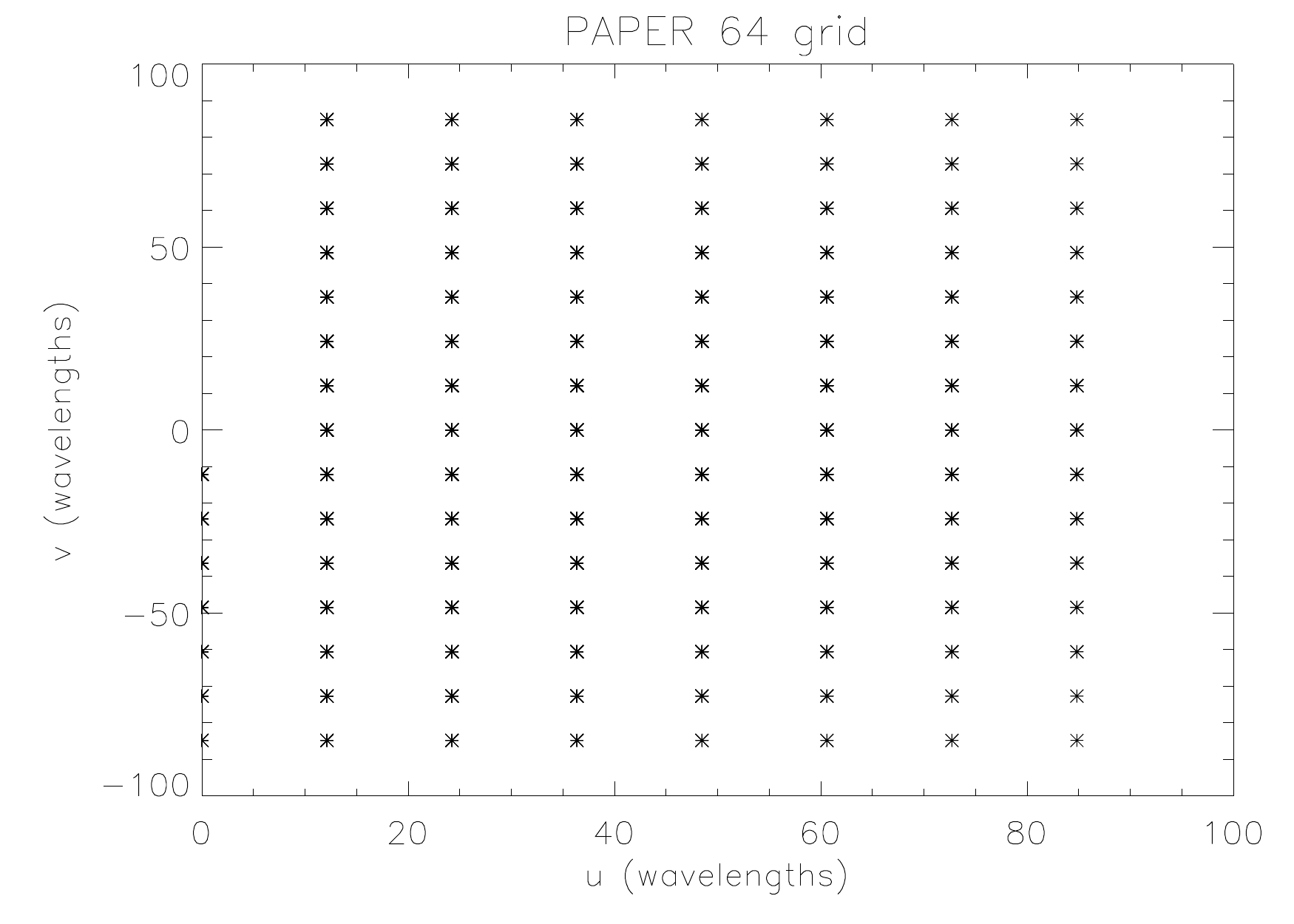}}
\vspace{1cm}
\caption{(Left) Antenna configurations for the MWA and PAPER considered in this work, and (right) instantaneous $uv$ coverage at zenith.}
\end{center}
\label{antenna_config}
\end{figure}

In addition to the proposed tile layout of the MWA, which is composed of a 1.5~km diameter core of 112 tiles and an outer region of 16 tiles (to the full 3~km maximum baseline), we will model a hypothetical instrument with an identical number of antennas and maximum baseline, but designed to provide uniform $uv$-coverage (approximately a Reuleaux triangle). We will also model PAPER, another low frequency instrument proposing to perform a statistical measurement of the EoR. PAPER currently comprises a 64-dipole instrument in a minimally-redundant configuration in the Karoo region of South Africa, with a maximum baseline of $\sim$260~m. As well as the current design, \citet{parsons11} discussed the advantages of building maximally-redundant instruments for EoR estimation, and one of their test designs will also be investigated in this work; a square 8$\times$8 grid with antenna separation of 26.3~m (yielding the same maximum baseline for both configurations). Figure \ref{antenna_config} displays the antenna configurations for the four arrays considered, and example instantaneous $uv$ coverage at zenith.
For all experiments we simulate an EoR field at a declination of $-30^o$, observing the field over a six hour window per day (hour angle range [-3,3] hours).

\subsection{EoR statistical estimation: angular power spectra}\label{power_spectrum_section}
The primary tools of a statistical measurement of the EoR are power spectra: angular power spectrum, two-dimensional [2D] power spectrum, and spherically-averaged power spectrum. These metrics quantify the signal power on a given angular or line-of-sight scale by forming a coherent$+$incoherent averaging of visibilities. We will consider the angular power spectrum and 2D power spectrum in this work.

The angular power spectrum is defined by \citep{morales04,mcquinn06,bowman09,datta10}:
\begin{equation}
C_l = \frac{\displaystyle\sum_{(uv) \in l}N_{uv}|V_{uv}|^2}{\displaystyle\sum_{(uv) \in l}N_{uv}},
\end{equation}
where $N_{uv}$ is the number of visibilities contributing to a given ($uv$) cell, and the sum is over the ($uv$)-cells contributing to that $l$-mode ($l=2\pi{|\boldsymbol{u}|}$). The visibilities, $V_{uv}$, are \textit{coherently} combined (averaged) within each ($uv$) cell, while the final estimate associating ($uv$)-cells in $l$-modes is \textit{incoherent} (visibilities are combined after squaring). Weighting by the number of contributing visibilities forms the maximum-likelihood estimate.

The 2D power spectrum is given by:
\begin{equation}
P(k_\bot, k_\parallel) = \frac{\displaystyle\sum_{|\boldsymbol{k}_\bot{|}\in k_\bot}W(\boldsymbol{k}_\bot, \boldsymbol{k}_\parallel)\left|V(\boldsymbol{k}_\bot, \boldsymbol{k}_\parallel)\right|^2}{\displaystyle\sum_{|\boldsymbol{k}_\bot{|}\in k_\bot}W(\boldsymbol{k}_\bot, \boldsymbol{k}_\parallel)},
\end{equation}
where the Fourier modes are in units of inverse distance (Mpc$^{-1}$), and are given by \citep[][and references therein]{peebles93,morales04}:
\begin{eqnarray}
k_\bot &=& \frac{2\pi{|\boldsymbol{u}|}}{D_M(z)},\\
k_\parallel &=& \frac{2\pi{H_0}f_{21}E(z)}{c(1+z)^2}\eta,
\end{eqnarray}
where $D_M(z), H_0, f_{21}, z$ are the transverse comoving distance, Hubble constant, frequency of the hyperfine transition ($\sim$1420~MHz), and observation redshift, respectively. $E(z)$ is \citep{hogg99}:
\begin{equation}
E(z) \equiv \sqrt{\Omega_M(1+z)^2 + \Omega_k(1+z)^2 + \Omega_\Lambda},
\end{equation}
and we assume a concordance cosmology \citep[$H_0$=70.4 kms$^{-1}$Mpc$^{-1}$, $\Omega_M$=0.27, $\Omega_k$=0, $\Omega_\Lambda$=0.73,][]{komatsu11}, and an observation redshift of $z$=8.

In practise, the power spectra are formed from an image cube of real quantities. In this work, we neglect the effects of gridding and image formation, and form the power spectra directly from the measured visibilities. This allows a clean propagation of errors from the measured data (i.e., the visibilities) to the final statistical measurement (i.e., the power spectrum). Because this involves (at most) one Fourier transform (FT, $\nu \rightarrow \eta\propto{k_\parallel}$), we need to be careful to simulate the usual three-dimensional FT ($[\theta_x,\theta_y,\nu] \rightarrow [u,v,\eta]$) from image space. A three-dimensional FT from real-valued to complex-valued quantities (as occurs transforming from the image plane) transforms from real to complex values. We typically choose the frequency direction as the final FT. In general, a real-to-complex FT of dimensions $N_XN_YN_Z$ transform to a complex cube of dimensions $N_XN_Y{\times}N_Z/2$. The final dimension contains $N_Z/2$ redundant channels, in the same way as a 1D real-to-complex FT. The choice of the redundant axis is arbitrary, but for convenience we choose the frequency direction. Under this scheme, both visibilities and their conjugates (Hermitian reflection in $uv$-plane) must be included in the power spectrum estimation, and only the non-redundant half of the $\eta$ array is used.


Forming the power yields a non-zero expected noise power ($\propto\sigma_V^2$). As discussed by \citet{mcquinn06} and \citet{bowman09}, the expected thermal power for proposed experiments will exceed the expected 21~cm EoR signal power. However, because the noise is stochastic, the expected power can be estimated and subtracted, or the experiment divided into two and cross-correlated, to eliminate the signal. These techniques cannot remove all of the noise power, because they are limited by sample variance, and power will remain at a lower level that corresponds to the \textit{uncertainty} in the magnitude of the noise component. The cross-correlation power spectrum is given by \citep{mcquinn06,bowman09,datta10}:
\begin{equation}
C_l = \frac{\displaystyle\sum_{(uv) \in l}N_{uv}|V_1^\ast{V_2} + V_2^\ast{V_1}|}{\displaystyle\sum_{(uv) \in l}2N_{uv}},
\end{equation}
where $V_1$, $V_2$ are the coherently-summed visibilities from each dataset. The noise variance (sample variance) remains after the expected power is removed. This process takes a coherent addition and replaces it with a partly incoherent addition, yielding a factor of two increase in the noise uncertainty.

Similar techniques could also be used to reduce the impact of the residual point source signal, assuming the residuals are uncorrelated. If there is a substantial degree of correlated noise between different angular modes, this technique will have limited use because it cannot remove correlated power, and structure will persist. Thus, it is the noise \textit{uncertainty} that is important for comparison. We demonstrate later that imprecise subtraction does yield correlated signal in both angular and spectral space, and this technique is therefore of limited use.

\section{Analysis}\label{analysis}
To determine the magnitude of residual signal from point source subtraction in the power spectra, we use both an analytic propagation of errors, and Monte-Carlo simulations. Ideally, would like to determine the full covariant uncertainties for all results, but the computational demands of forming large, complex-valued covariance matrices necessitate the use of simulations for the two-dimensional power spectrum.

In this section we begin by describing the construction of the power spectra calculations (\ref{construction}). We then derive the source position uncertainties based on the CRB analysis (\ref{uncert_cal}) and the covariant propagation of these uncertainties to visibilities (section \ref{error_prop_section}). These are then used as starting points for both the analytic error propagation (\ref{error_prop_section2}, used for the one-dimensional power spectrum), and for the simulations (\ref{simulations}, used for the two-dimensional power spectrum).

\subsection{Construction of estimates}\label{construction}
In a real EoR experiment, the $uv$-gridding scale (cell size) is the region over which the visibilities can be considered coherent, and is determined by the instrument field-of-view and desired sampling of the synthesized beam (yielding a pixel scale in image space). For the MWA, and sampling the beam with five pixels, this corresponds to $\sim$1000$\times$1000 $uv$ cells. The power in each of these cells is then combined incoherently to form the power spectrum estimate for a given angular scale. It is computationally infeasible to perform a covariant analysis with a complex matrix with $\sim$10$^{12}$ entries. For the purposes of this work, where we seek (1) to determine the magnitude of the residual point source signal relative to the thermal noise component, and (2) to determine the level of correlation between angular modes via a fully covariant analysis, we use a very coarse sampling of the $uv$ plane to form the intermediate, coherent $uv$ cells (a 40$\times$40 grid of cells, yielding a 1600$\times$1600 complex covariance matrix) for our covariant analytic computation. This yields an estimate of the angular power that is more coherent than is possible with a real experiment, effectively coherently combining visibilities that are not coherent in practise. In addition, we coarsely bin the angular scales ($l$-modes) for computational ease.

Appendix A contains a derivation of the expected power and power variance for a given coherent/incoherent combination of visibilities. This is useful both for an understanding of how to design a real EoR experiment, and also in the context of the computational simplifications we have employed in the analytic covariant calculation. A further consequence of this simplification is that the estimated magnitude of the uncertainty (square-root of the sample variance) of the thermal noise power, and of the residual point source subtraction signal power will be comparable to the expected power; this directly follows from the unphysically-coherent calculation we are performing.

In addition to the covariant angular power spectrum calculations, we perform an analytic variant 1D calculation with $(uv)$ sampling appropriate for a real experiment. This yields realistic results because we expect the angular modes to be uncorrelated due to Earth rotation synthesis over the contributing baselines (structure will be suppressed by the rotation of the array baselines with respect to the observation field). This analysis is not possible for the 2D angular power spectrum, where frequency-based correlations are known to be important for reproducing realistic results, and indeed are the basis for the ``wedge" feature observed in simulations in previously published results. For the 2D power spectrum, we use Monte-Carlo simulation.

\subsection{Calculation of uncertainties in calibrator parameters}\label{uncert_cal}
\begin{figure}
\begin{center}
\includegraphics[scale=0.75]{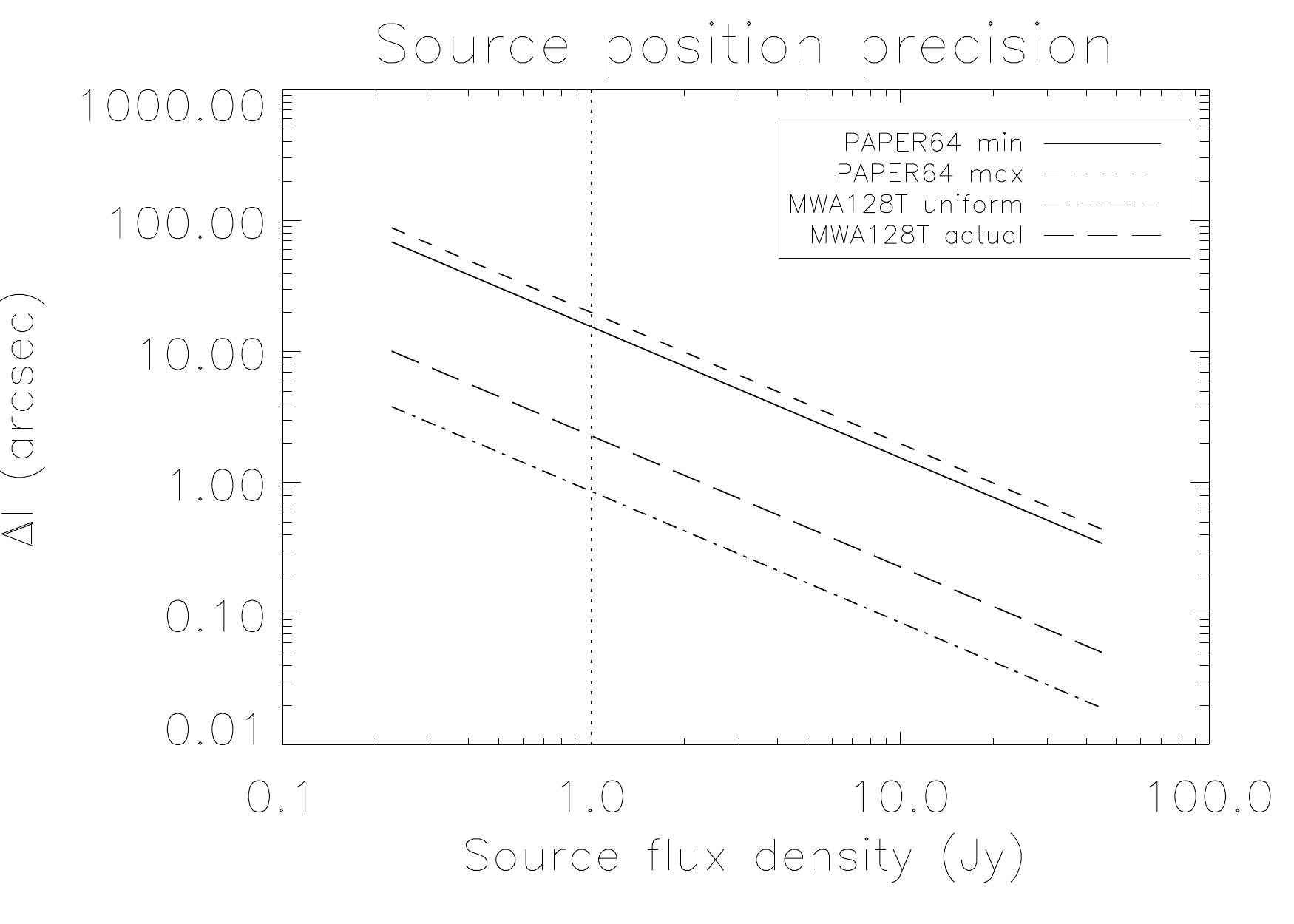}
\caption{Optimal calibrator position precision (arcseconds) versus source flux density (Jy) for the instruments considered in this work, and using the full instantaneous bandwidth of the instrument (Table \ref{instrument_design}: BW = 30.72~MHz (MWA), 70.0~MHz (PAPER); $\Delta{t}$=8~s, T$_{\rm sys}$=440~K). The vertical line denotes the minimum flux density for peeling (1~Jy).}\label{source_precision_plot}
\end{center}
\end{figure}

The signal in each visibility is the linear combination of signals from each calibrator, and is given by:
\begin{equation}
\tilde{s}[f,n] = \displaystyle\sum_{i=1}^{N_c} V_i(u_{fn},v_{fn}) = \displaystyle\sum_{i=1}^{N_c} B_i(l_i,m_i)\exp{\left[-2{\pi}i(u_{fn}l_i+v_{fn}m_i)\right]},
\label{visibility_signal}
\end{equation}
where $N_c$ is the total number of calibrators, described by source strength, $B_i$, located at sky position, ($l_i,m_i$), for a baseline, $n$, and frequency channel, $f$ (we have assumed flat-spectrum calibrators throughout). The signal is embedded within white Gaussian thermal noise, with diagonal covariance matrix, $\boldsymbol{C}=\sigma^2\boldsymbol{I}$, where $\sigma$ is set by the correlator integration time ($dt$) and channel width ($d\nu$), and is given in temperature units by (single polarization):
\begin{equation}
\sigma = \frac{A_eT_{\rm sys}}{\lambda^2\sqrt{d\nu{dt}}} \hspace{0.5cm} [{\rm K}],
\end{equation}
where $A_e$ is the effective antenna area at wavelength, $\lambda$, and $T_{\rm sys}$ is the system temperature.
Using the full dataset, the minimum uncertainty in the parameter estimates for calibrator, $i$, and their non-zero covariances, are given by \citep{trott11}:
\begin{eqnarray}
\Delta{l_i} &\geq& \frac{\sigma I_{v^2}^{1/2}}{2\sqrt{2}\pi{B_i}} 
\left[ I_{v^2} I_{u^2} - \left(I_{uv} \right)^2 
\right]^{-1/2} \label{l_precision}\\
\Delta{m_i} &\geq& \frac{\sigma I_{u^2}^{1/2}}{2\sqrt{2}\pi{B_i}} 
\left[ I_{v^2} I_{u^2} - \left(I_{uv}\right)^2 \right]^{-1/2} \label{m_precision}\\
\Delta{B_i} &\geq& \frac{\sigma}{\sqrt{2NF}} \\ 
{\rm cov}(l_i,m_i) &=& -\frac{\sigma^2 I_{uv}}{8\pi^2{B_i}^2} 
\left[ I_{v^2} I_{u^2} - \left(I_{uv}\right)^2 \right]^{-1}
\end{eqnarray}
 where
\begin{eqnarray}
&I_{u^2}& = \displaystyle\sum_{n=1}^N \displaystyle\sum_{f=1}^F  u_{fn}^2,  \\
&I_{uv}& = \displaystyle\sum_{n=1}^N \displaystyle\sum_{f=1}^F  u_{fn}v_{fn},  \\
&I_{v^2}& = \displaystyle\sum_{n=1}^N \displaystyle\sum_{f=1}^F v_{fn}^2, \label{Iv2}
\end{eqnarray}
and $N$ is the number of baselines, $F$ is the number of frequency channels.
Figure \ref{source_precision_plot} plots the optimal point source position precision ($\Delta{l}$) as a function of calibrator signal strength (Jy) for the four instruments considered.
The linear dependence of precision with signal strength is observed. The inner core $+$ outer ring structure of the MWA degrades its performance slightly compared with the uniform array. The short baselines and fewer antennas of the PAPER arrays degrade their ability to localise sources well compared with the MWA, although their wider instantaneous bandwidth balances this effect somewhat. The vertical line at 1~Jy displays the peeling cutoff considered in this work.

For ionospheric calibration, each calibrator source provides information about the amplitude and direction of the ionospheric phase shift, by comparison with a sky model of the true calibrator location. Both the overall and differential refraction are important for a full ionospheric model, but the overall shift can be estimated jointly from the calibrators, while the local refraction is independent (in the simplest case where local correlations are ignored). For faint calibrator sources, the precision theoretically achievable with an 8 second dataset may be comparable to the differential refraction due to the ionosphere. In this work, we assume no \textit{a priori} information about source location from previous calibration solutions, and use the information contained within the current visibilities alone to estimate source precision. This a reasonable assumption for the MWA, but not necessarily for faint sources with PAPER. For this reason, we limit the minimum peeling flux density to 1~Jy in this work. In \citet{datta10}, because all sources are assumed to have the same RMS position precision, the brightest source in the field contributes the greatest residual error. They concluded that 0.1 arcsec precision was required. Figure \ref{source_precision_plot} shows that this precision can, in principle, be achieved by the MWA with an 8~s snapshot for sources brighter than 30~Jy. Sources fainter than this cannot be subtracted with this precision, however the magnitude of the error in this case is lower due to the source's lower flux. In the next section we demonstrate how sources in the field contribute to the overall residual error.

For the experiments considered in this work, we use the full bandwidth of the instrument (Table \ref{instrument_design}) to determine the precision of source parameter estimates, regardless of the bandwidth used for the individual power spectra. This provides the highest precision on parameter estimates available to the instrument.

\subsection{Propagation of errors into interferometric visibilities}
\label{error_prop_section}

\begin{figure}
\begin{center}
\includegraphics[scale=0.55]{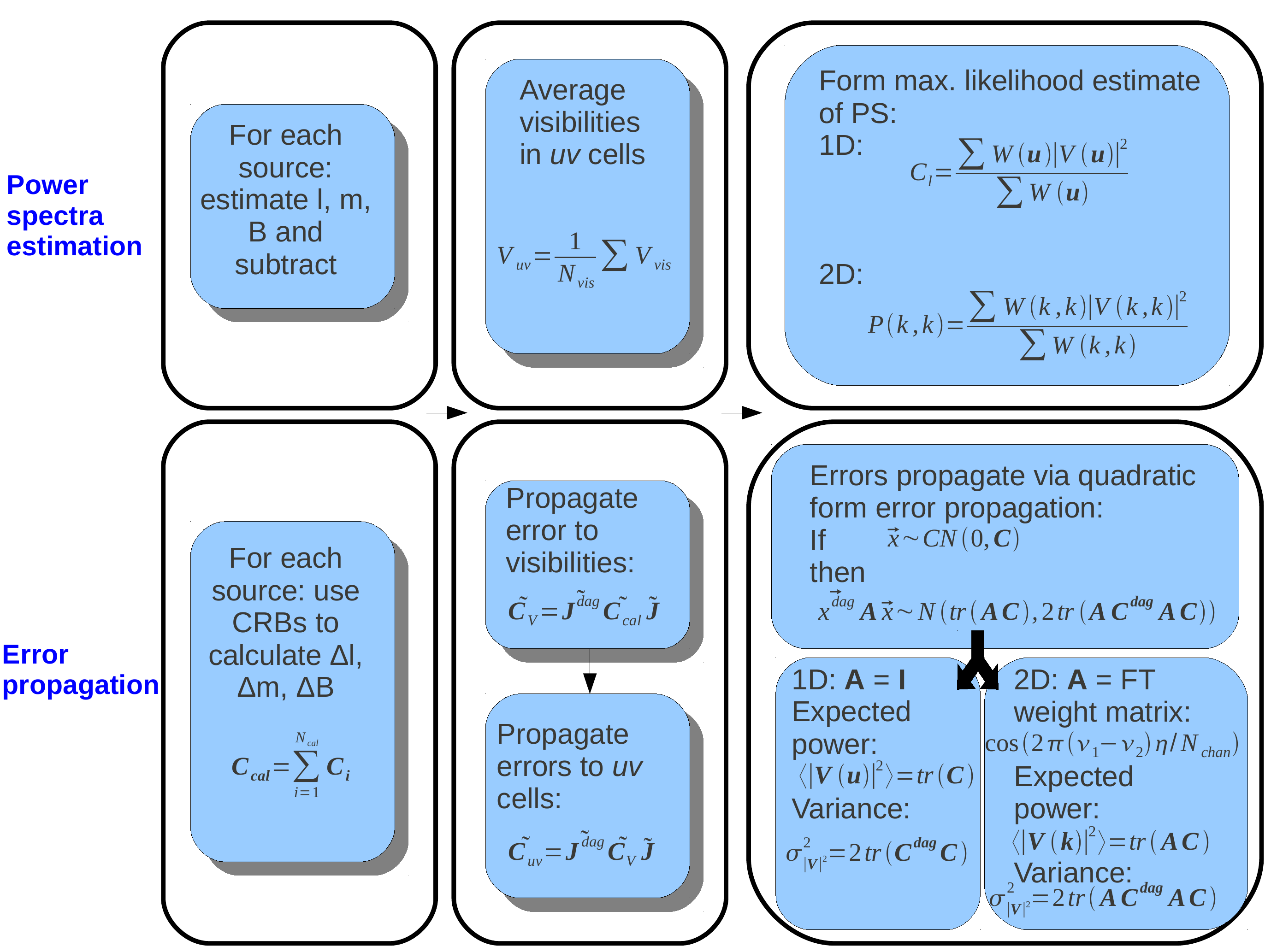}
\caption{Diagram showing the steps in estimating the power spectrum (upper boxes), and errors on the power spectrum (lower boxes).}\label{methods_diagram}
\end{center}
\end{figure}

Figure \ref{methods_diagram} displays schematically the steps involved in forming the power spectrum, and in propagating the errors from residual point source signal to the power spectrum.

The uncertainties in signal parameters for each calibrator are propagated into uncertainty in the measured visibilities, in addition to the statistical thermal noise. Because the measurement errors for each calibrator are independent, the overall visibility-based noise covariance matrix will be the sum of the matrices for each calibrator. The error propagation follows a fully-covariant treatment, where the residual signal noise covariance for calibrator, $i$, is given by:
\begin{equation}
\boldsymbol{C}_{V_i} = \boldsymbol{J}\boldsymbol{C}_{\theta_i}\boldsymbol{J}^\dagger,
\label{jacobian}
\end{equation}
where $\boldsymbol{J}$ is the $3\times{N_{\rm vis}}$ complex Jacobian of partial derivatives of the visibility functions with respect to the parameters ($\theta_i=(l_i,m_i,B_i)$), $\boldsymbol{C}_{\theta_i}$ is the $3\times{3}$ covariance matrix of parameter uncertainties, and the dagger denotes the complex-conjugate transpose operation.

The treatment provides a full covariant analysis, including any correlations between visibilities due to residual noise power. For a given visibility and a single calibrator, the variance is given by (using equation \ref{jacobian}):
\small
\begin{eqnarray}
\sigma_V^2 &=& \left|\frac{\partial{V}}{\partial{l}} \right|^2 \sigma_l^2 + \left|\frac{\partial{V}}{\partial{m}} \right|^2 \sigma_m^2 + 2\left(\frac{\partial{V}^\ast}{\partial{l}}\frac{\partial{V}}{\partial{m}}+\frac{\partial{V}}{\partial{l}}\frac{\partial{V}^\ast}{\partial{m}} \right) {\rm cov}_{lm} + \left|\frac{\partial{V}}{\partial{B}} \right|^2 \sigma_B^2\label{error_vis1}\\
&=& 4\pi^2u^2B^2\frac{\sigma^2}{8\pi^2B^2}f_1(u,v) + 4\pi^2v^2B^2\frac{\sigma^2}{8\pi^2B^2}f_2(u,v) + 8\pi^2uvB^2\frac{\sigma^2}{8\pi^2B^2}f_3(u,v) + \frac{\sigma^2}{2N_{\rm vis}}\\
&\propto& u^2f_1(u,v) + v^2f_2(u,v) + 2uvf_3(u,v) + \frac{1}{2N_{\rm vis}},\label{error_vis3}
\end{eqnarray}
\normalsize
where $f_1,f_2,f_3$ are functions of the baseline lengths (array geometry). The residual error in each visibility due to each calibrator is \textit{independent} of the calibrator signal strength. Therefore, the impact on the visibilities of subtracting each calibrator has the same magnitude for all calibrators [but different distributions across visibilities; the covariances also depend on the source positions, ($l_i,m_i$)]. This result deviates from the assumptions of previous work, which assumed that the source position error was independent of signal strength, and therefore that residual error was greatest for the strongest calibrators.

Because the residual signal is due to a real source (signal-like), and not a random process (noise-like), the errors will be correlated between baselines and between frequency channels, for a given estimation of the source parameters. These covariances yield a correlated noise covariance matrix for the visibilities, deviating from the uncorrelated white noise of the stochastic thermal component (diagonal covariance matrix). Such covariances propagate from the visibilities to the angular power spectra, potentially yielding correlations between angular modes.

\subsection{Propagation of errors into angular power spectra}\label{error_prop_section2}
As described schematically in Figure \ref{methods_diagram} and section \ref{power_spectrum_section}, the high-level steps to calculating the angular power spectrum from visibility data are \citep[][and references therein]{mcquinn06}:
\begin{itemize}
\item Visibilities are averaged into pre-defined $uv$-cells in each frequency channel, yielding a coherent combination of information;
\item For multiple frequency channels (two-dimensional power spectrum), averaged visibilities are Fourier transformed in frequency to form the line-of-sight power ($\nu$--$\eta$ Fourier pair);
\item Averaged complex visibilities are squared to form real-valued power estimates in each $uv\eta$-cell;
\item The maximum likelihood estimate of the power spectrum is computed by forming a visibility-weighted power estimate in each angular and line-of-sight mode ($k_\bot$, $k_\parallel$), from that mode's contributing $uv\eta$ cells.
\end{itemize}
The final step can be varied in two or three dimensions to yield one- and two-dimensional power spectra estimates.

The corresponding error propagation from visibilities to angular power spectra follows a fully-covariant error propagation, given broadly by:
\begin{itemize}
\item Propagate errors from visibilities to $uv$ cells according to $\boldsymbol{C}_{uv} = \boldsymbol{J}\boldsymbol{C}_{V_i}\boldsymbol{J}^\dagger$, where the Jacobian contains normalizations for the number of visibilities contributing to each cell, and zeroes otherwise;
\item Squaring averaged visibilities to form power estimates yields a quadratic form (i.e., a metric quadratic in the data). The Fourier transform from observing frequency ($\nu$) to line-of-sight frequency ($\eta$) can be incorporated into the quadratic form via a weighting matrix. The errors for one- and two-dimensional power spectra propagate according to \citep{searle1971};
\begin{equation}
{\rm If} \hspace{0.1cm} \boldsymbol{x} \sim CN(0,\boldsymbol{C}) \hspace{0.2cm} {\rm then} \hspace{0.2cm} \boldsymbol{x^{\dagger}Ax} \sim N[{\rm tr}(\boldsymbol{AC}),2{\rm tr}(\boldsymbol{AC^{\dagger}AC})]
\label{quadratic_form_eqn}
\end{equation}
where $\boldsymbol{A}$ is a real weighting matrix, and is given by,
\begin{itemize}
\item 1D: $\boldsymbol{A}=\boldsymbol{I}$;
\item 2D: ${A}_{\alpha\beta}=\cos{(2\pi(\nu_\alpha-\nu_\beta)\eta/N)}$, where $N$ is the number of frequency channels.
\end{itemize}
\item Propagate to the incoherent addition of power samples (the maximum likelihood estimate of the power spectrum) via the error propagation, $\boldsymbol{C}_{P} = \boldsymbol{J}\boldsymbol{C}_{||^2}\boldsymbol{J}^\dagger$, where the Jacobian entries contain the number of visibilities contributing to each ($k_\bot$, $k_\parallel$) cell.
\end{itemize}
After cross-correlation to remove persistent noise, the covariances calculated according to equation \ref{quadratic_form_eqn} form the noise variance that would contaminate the power spectrum (see discussion in Section \ref{power_spectrum_section}).

\subsection{Monte-Carlo simulations}\label{simulations}
As discussed in section \ref{analysis}, the expected power and power uncertainty for the 2D power spectrum is obtained via Monte-Carlo simulation. To produce realistic results that are consistent with our theoretical framework, we generate our simulated visibilities from the correlated CRB covariance matrix of source parameter uncertainties (including source position covariances, equations \ref{l_precision}--\ref{Iv2}). That is, for each antenna configuration, we simulate visibilities with thermal noise and residual point source signal (obtained by generating Gaussian random variables, with a correlation structure given by the CRB results) at the appropriate $uv$ co-ordinates for the given experiment.

We follow the standard methodology to form power spectra from visibility data (without transforming to the image plane), as described in section \ref{power_spectrum_section}. We simulate 100 noise realizations for each experiment, with parameters given by Table \ref{instrument_design}, and with $(uv)$ and ($k_\bot,k_\parallel$) sampling appropriate for a real experiment (e.g., 1000$\times$1000 $uv$ cells, and ten $k$ bins per decade, for MWA). In this way, we achieve realistic power spectra using the theoretical source position errors derived from the bounds.

\section{Results}
We have presented the formalism for deriving the expected power and power covariance in angular power spectra. Computing the covariances between angular or line-of-sight modes provides an understanding of the level of correlation in the data, introduced by the peeling process. In addition to desiring the amplitude of the peeling residuals, compared with thermal noise, we also require an understanding of the correlations between modes, to perform an optimal and unbiased estimate of EoR parameters from power spectra data. As described above, performing a fully-covariant analysis is limited by the computational requirements of calculating and storing very large matrices. Thus, we also perform variant analyses and Monte-Carlo simulations.

We now present the results of an \textit{analytic covariant} angular power spectrum, which is binned coarsely for computational feasibility and provides information about the level of correlation between $l$ modes. In addition to the covariant calculation, we then present an \textit{analytic variant} angular power spectrum, which is binned finely, and reflects the conditions available to real experiments (Section \ref{angular_pow_spec_section}). We then present the results of a \textit{simulated covariant} 2D power spectrum (where the covariances are carried naturally through the simulation, but without the number of simulations required to form an accurate estimate of the mode covariances), which is binned finely, and also reflects the conditions of a real experiment (Section \ref{2d_pow_spec_section}).

\subsection{Angular power spectrum}\label{angular_pow_spec_section}
\begin{figure}
\subfigure[MWA uniform $uv$ coverage.]{\includegraphics[scale=0.25]{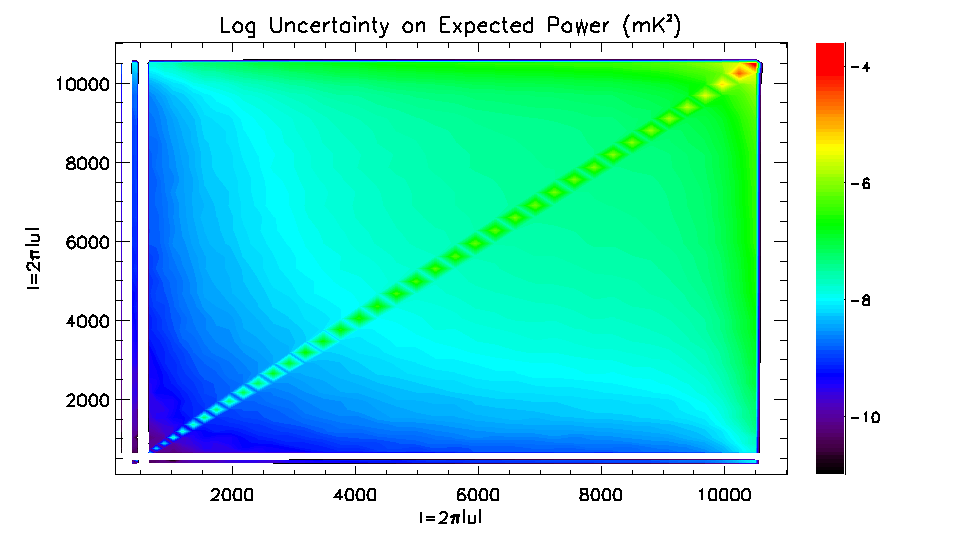}}
\subfigure[MWA.]{\includegraphics[scale=0.25]{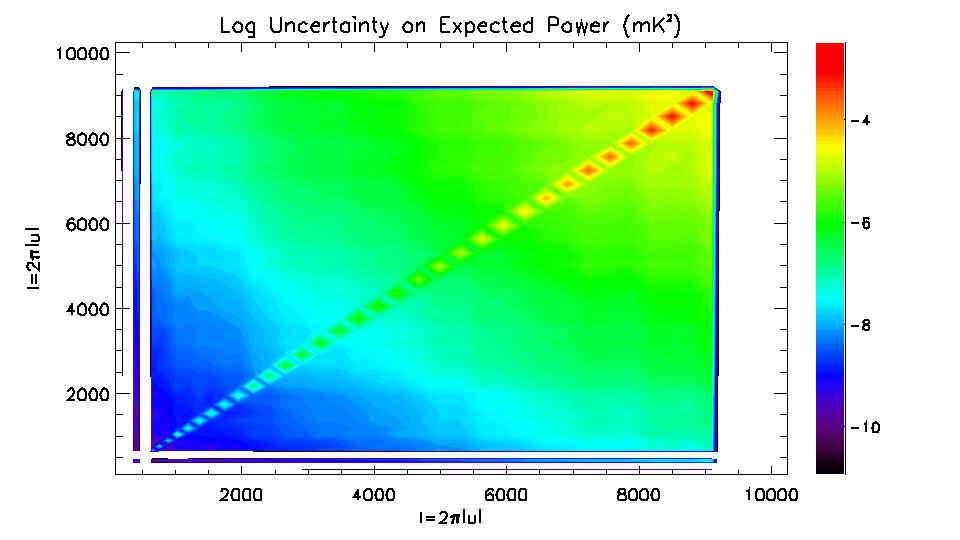}}\\
\subfigure[Actual 64-dipole minimum-redundancy (PAPER).]{\includegraphics[scale=0.25]{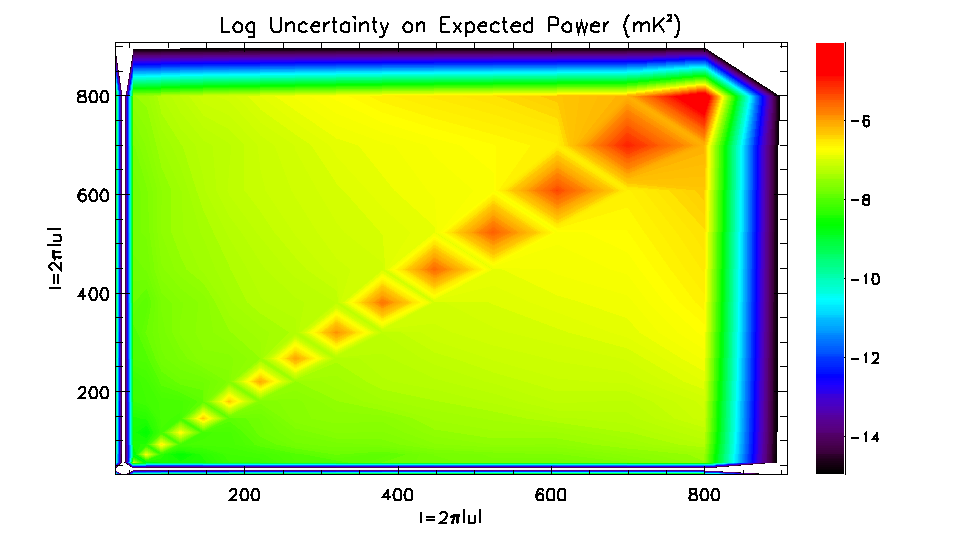}}
\subfigure[Potential 64-dipole maximum-redundancy (PAPER).]{\includegraphics[scale=0.25]{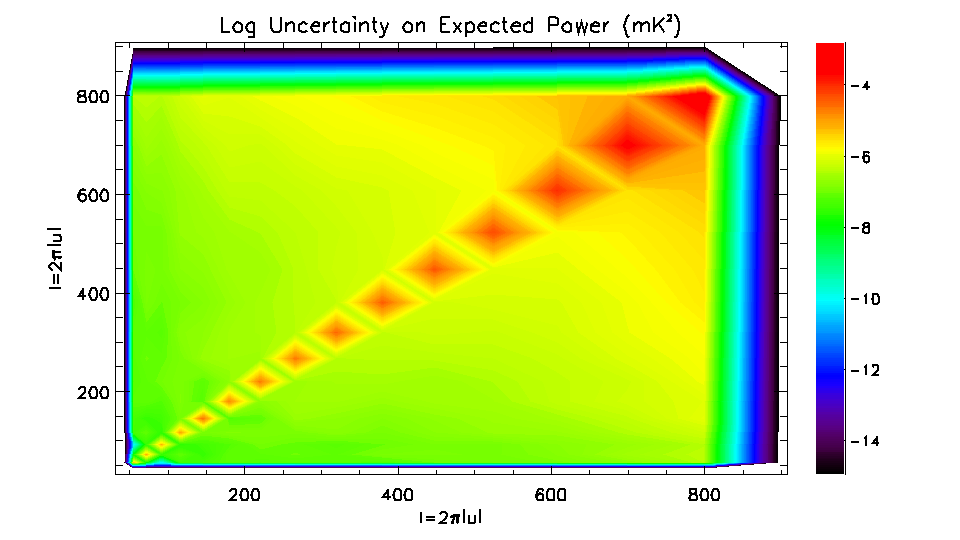}}
\vspace{1cm}
\caption{Square-root of covariance for angular power spectrum (mK$^2$).}
\label{1d_covariance}
\end{figure}

\begin{figure}
\subfigure[MWA uniform $uv$ coverage.]{\includegraphics[scale=0.45]{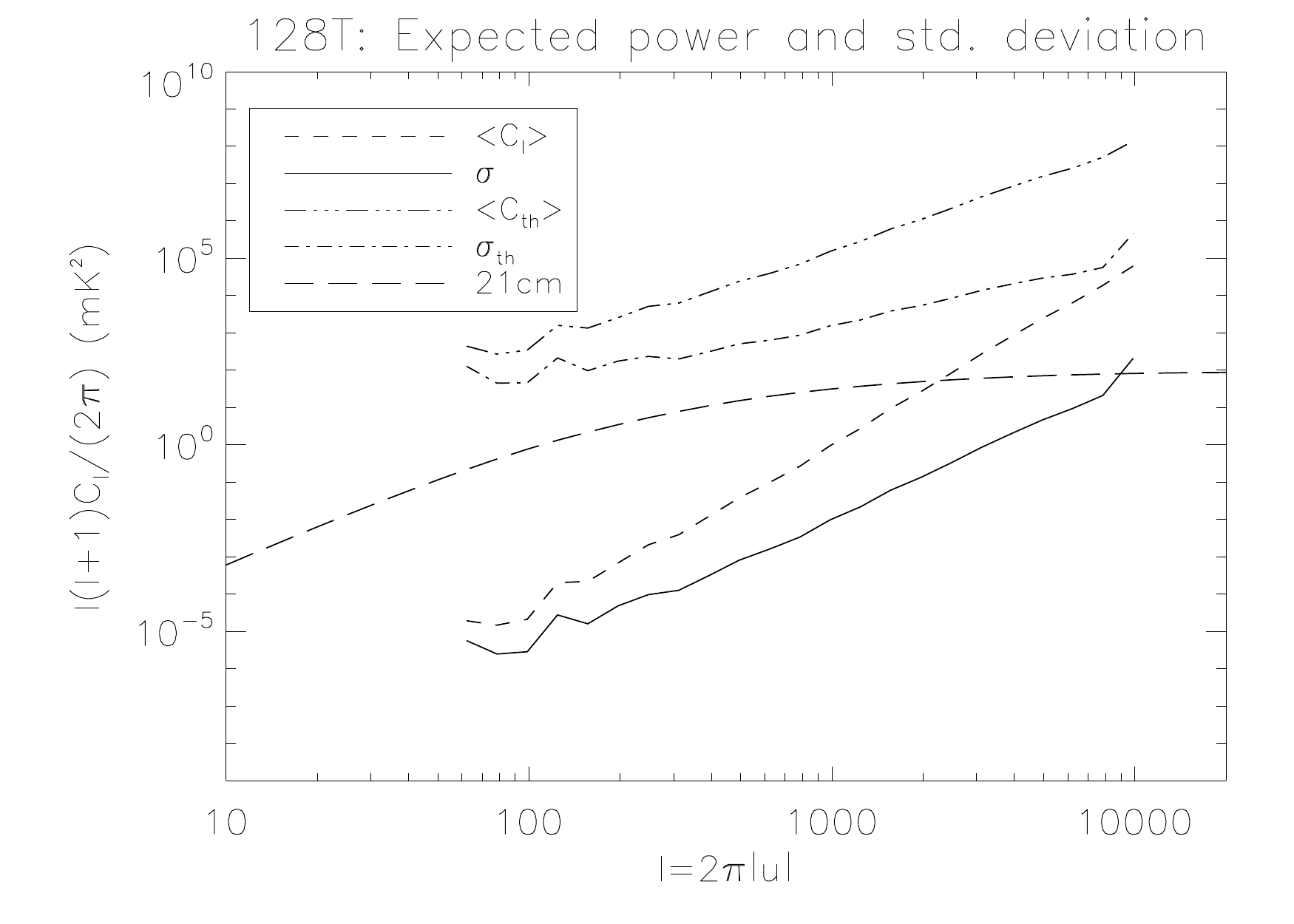}}
\subfigure[MWA.]{\includegraphics[scale=0.45]{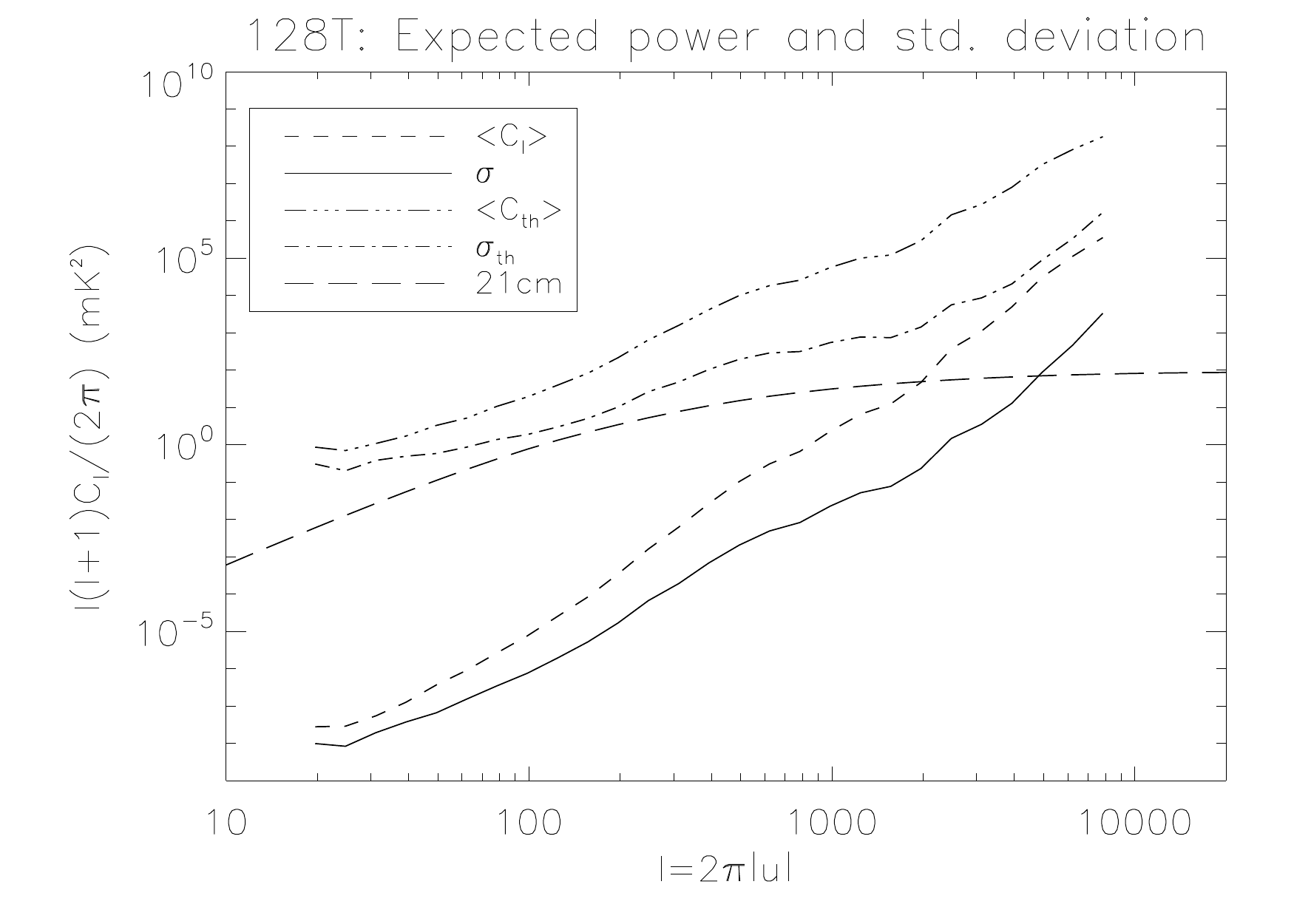}}\\
\subfigure[Actual 64-dipole minimum-redundancy (PAPER).]{\includegraphics[scale=0.45]{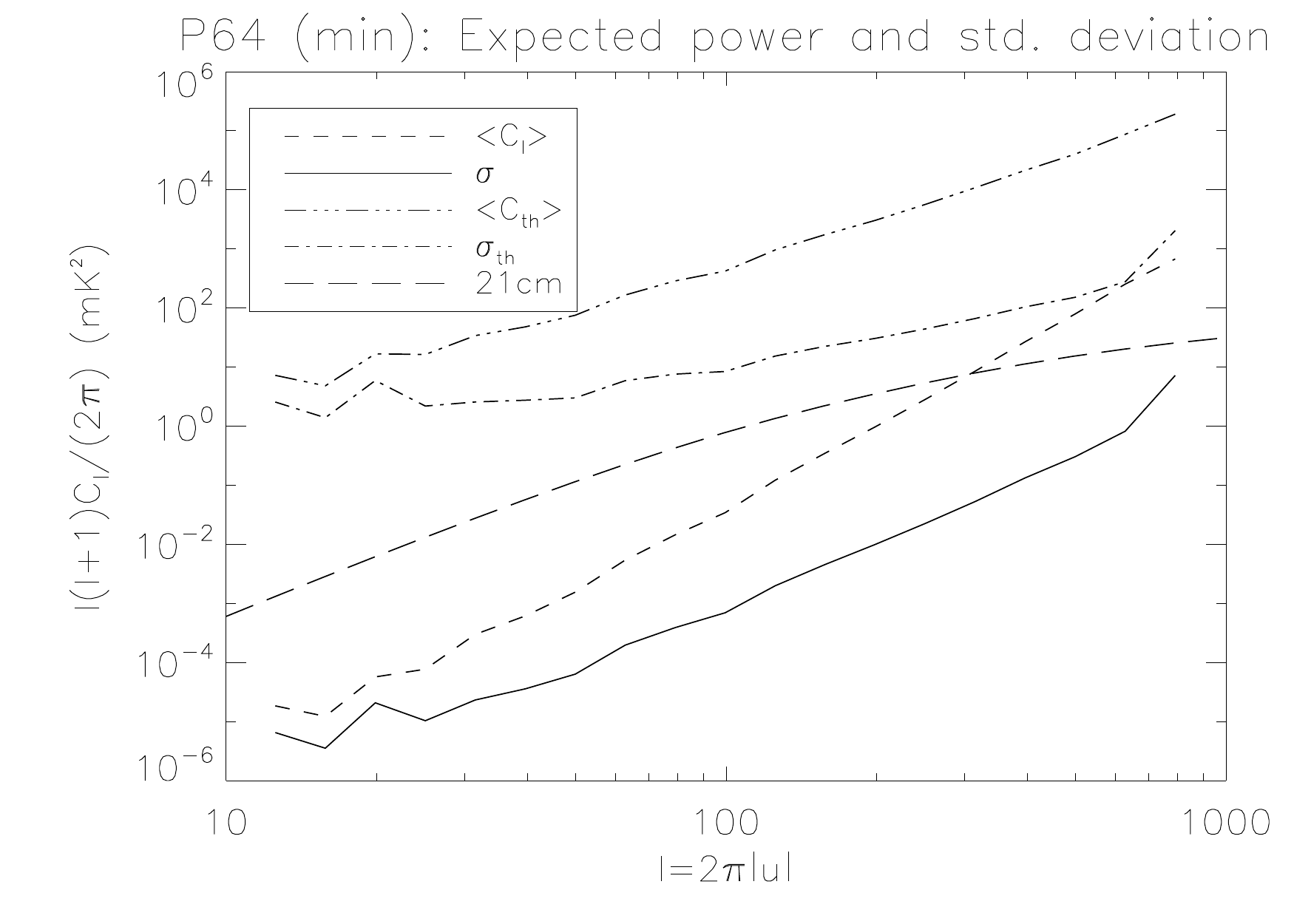}}
\subfigure[Potential 64-dipole maximum-redundancy (PAPER).]{\includegraphics[scale=0.45]{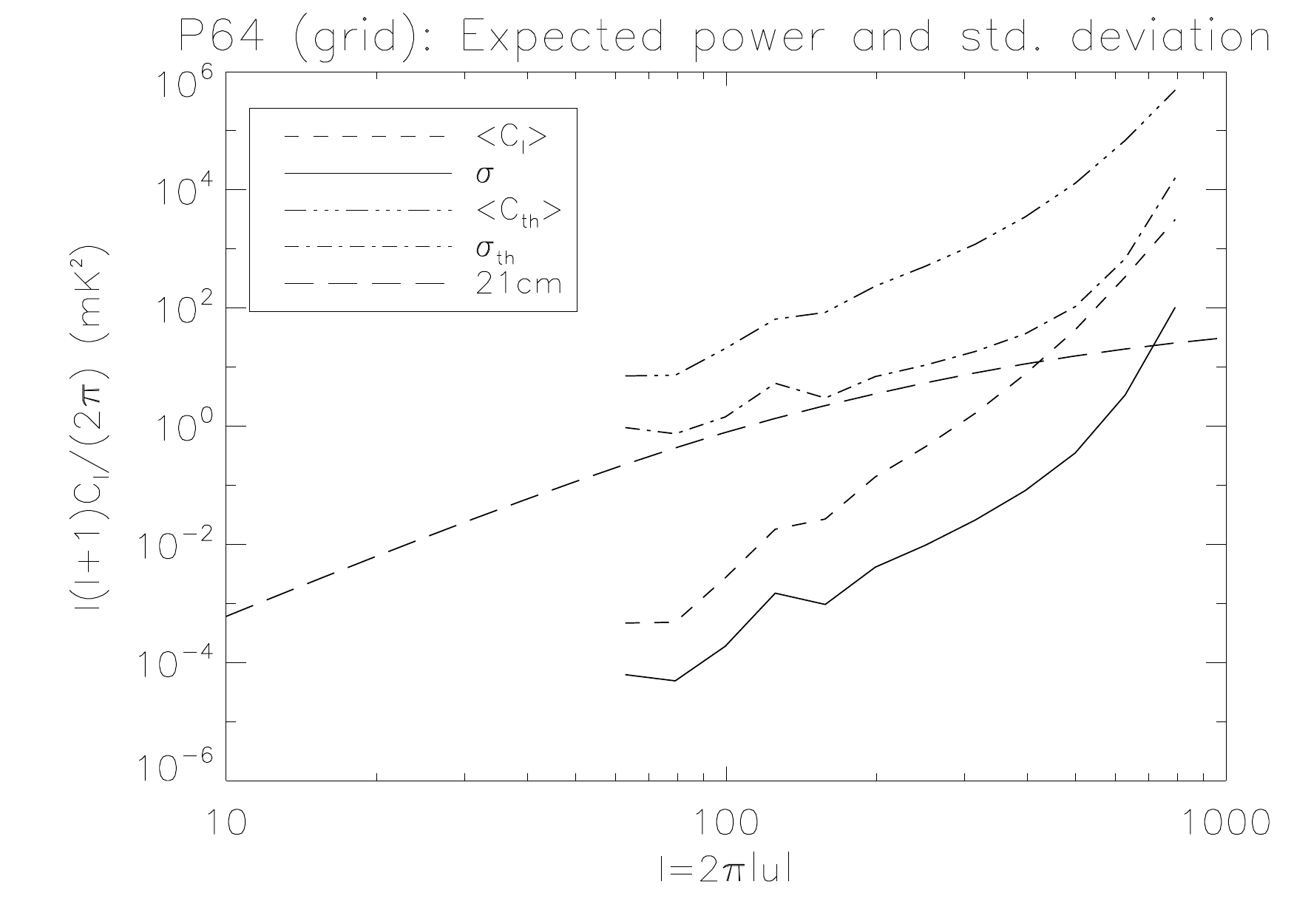}}
\vspace{1cm}
\caption{Expected power and uncertainty for thermal noise and residual signal (mK$^2$), using a variant calculation with realistic $(uv)$ cells. The binning is logarithmic with ten bins per decade in $l$. The expected contribution from the 21~cm signal is also shown for comparison.}
\label{1d_variance}
\end{figure}

Figures \ref{1d_covariance} and \ref{1d_variance} display the fully-covariant, and variant, analytic calculation results for the angular power spectrum, for each of the four experiments considered, and with experimental parameters defined in Table \ref{instrument_design}.

Figure \ref{1d_covariance} shows the \textit{uncertainty} on the residual signal power ($\sqrt{\boldsymbol{C}_{C_l}}$), given by the square-root of the covariance term in equation \ref{quadratic_form_eqn}. It plots $l-l$ and is symmetric about the diagonal, which represents the variance in each $l$-mode (note that no factor of $l(l+1)/2\pi$ has been included here). Covariances are $\sim$2 orders of magnitude below the variances, indicating that angular modes are not substantially correlated. This result is unsurprising given that we expect that Earth rotation synthesis of the field across six hours of sky will average out any correlations between baselines (note that rotation is not expected to remove frequency channel-based covariances, which rotate with the other visibilities on the same baseline). This result suggests that correlations between angular modes are minimal, and EoR parameters derived from measured power spectra will not be strongly biased by assuming uncorrelated data.

Figure \ref{1d_variance} displays the expected power ($l(l+1)C_l/2\pi$) and power uncertainty ($\sqrt{\sigma_{C_l}^2}$), for the thermal noise and residual signal, as well as the expected 21~cm signal (assuming a fully-neutral IGM at $z$=8, where we have used a linear power spectrum and assumed isotropy). In all cases the thermal noise power and uncertainty exceed the power from the point source subtraction. One can see the differences in the gradients of these noise components, with the higher angular modes suffering more severely from point source subtraction, consistent with the additional noise in the long baseline visibilities (equation \ref{error_vis3}). These figures also highlight the differences between the array configurations; the minimally-redundant arrays (Figure \ref{1d_variance}a,c) exhibit uniform behaviour across a wide range of angular modes, while the non-uniform arrays (Figure \ref{1d_variance}b,d) sample particular modes well, but have a reduced range of sampled $l$-modes. In particular, the maximally-redundant PAPER grid array displays slightly improved performance at the lowest sampled modes (effectively the grid spacing), while performing poorly at the higher modes, and sampling a reduced range of modes. The improvement at small $l$ is limited by the logarithmic binning of the $l$-modes, yielding fewer visibilities contributing than would be the case for a linear binning.

The MWA configurations show similar behaviour: the uniform array yields even performance across the angular scales, while the centrally-concentrated array samples smaller $l$ modes, yields improved performance at small $l$, and poorer performance on the longer baselines (incorporating the few antennas on the outer ring).

\subsection{2D power spectrum}\label{2d_pow_spec_section}
\begin{figure}
\subfigure[MWA uniform $uv$ coverage.]{\includegraphics[scale=0.27]{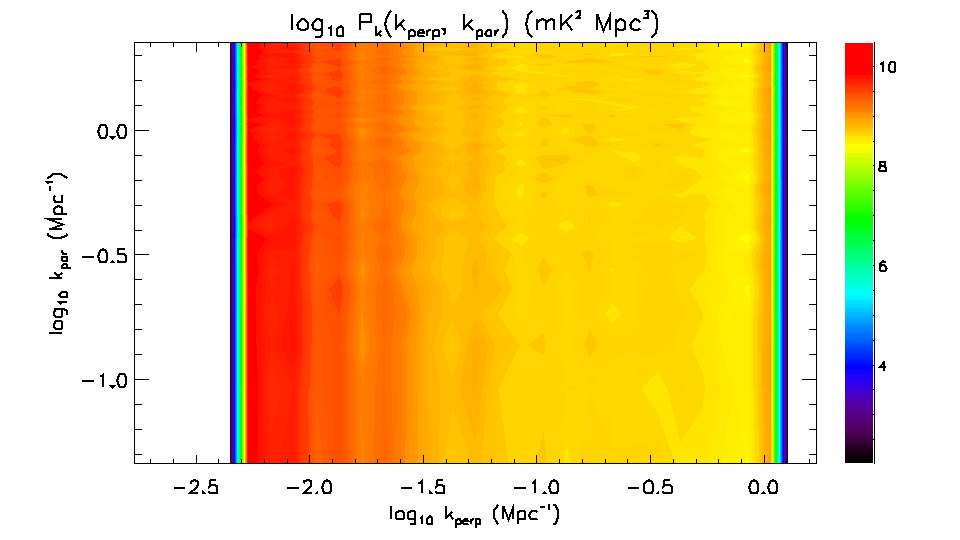}}
\subfigure[MWA.]{\includegraphics[scale=0.27]{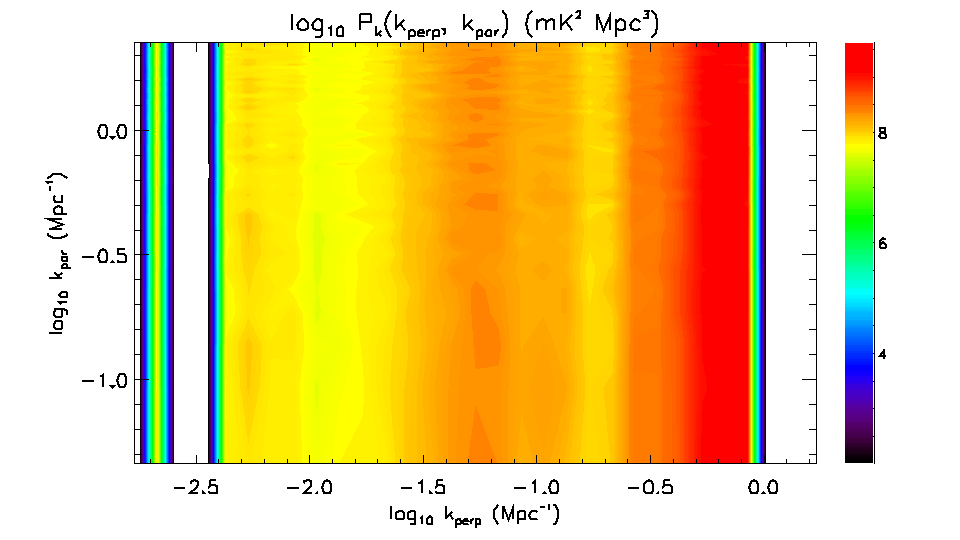}}\\
\subfigure[Actual 64-dipole minimum-redundancy (PAPER).]{\includegraphics[scale=0.27]{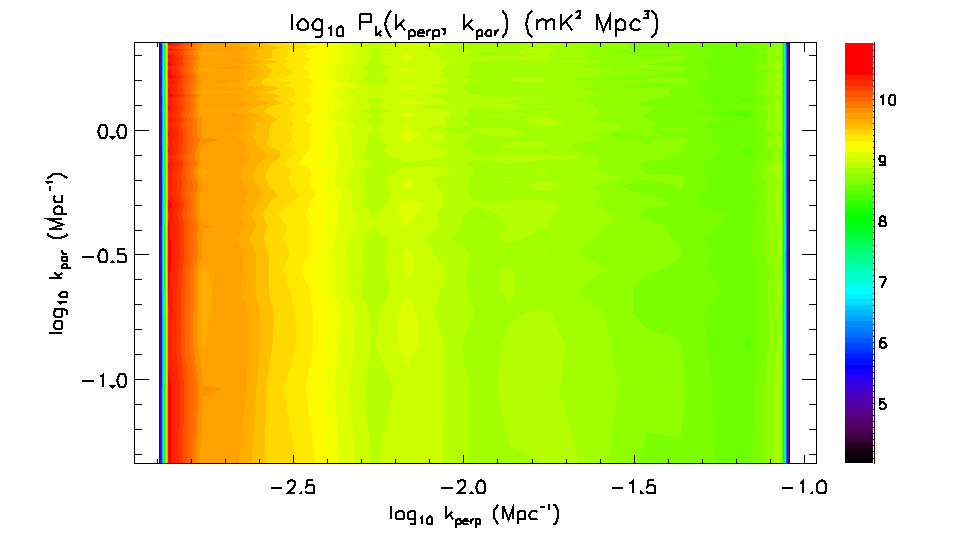}}
\subfigure[Potential 64-dipole maximum-redundancy (PAPER).]{\includegraphics[scale=0.27]{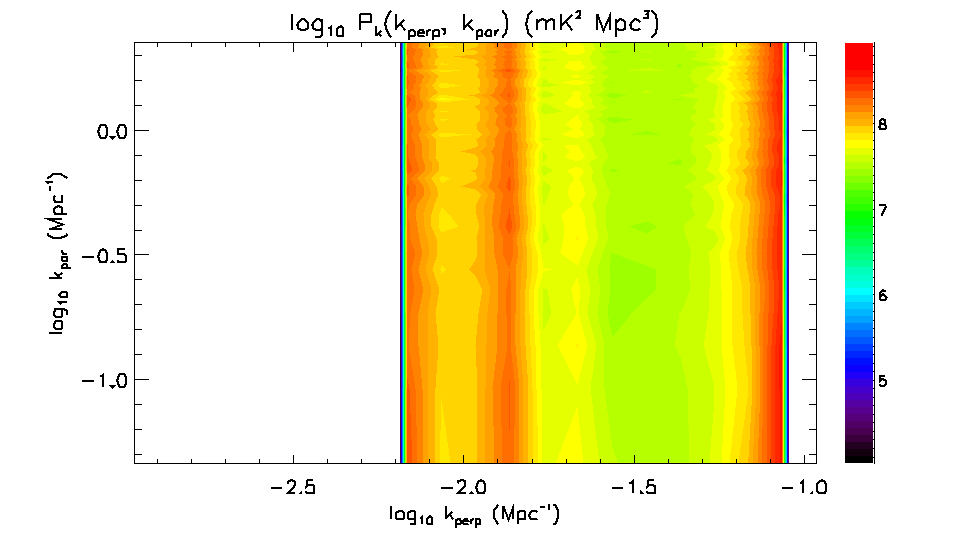}}\\
\subfigure[Expected 21~cm HI signal.]{\includegraphics[scale=0.27]{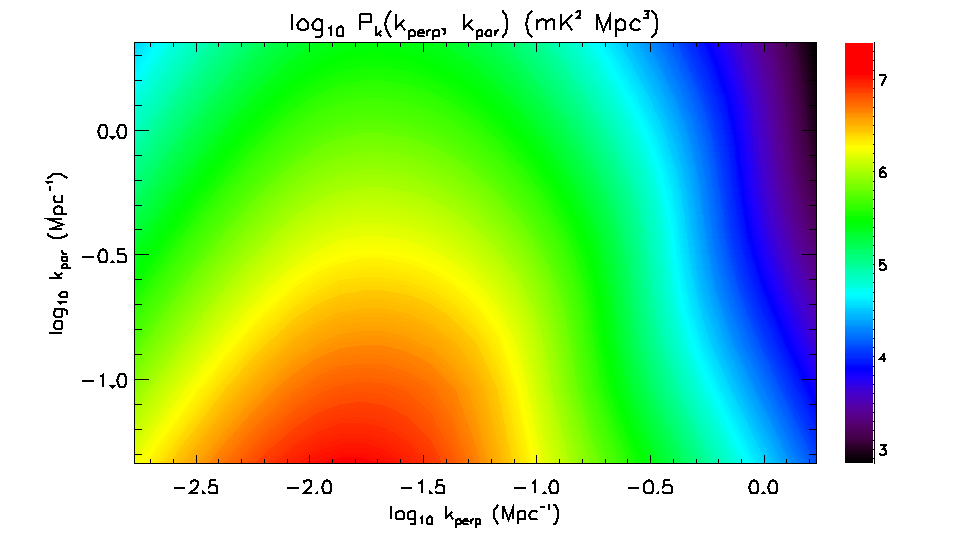}}
\vspace{1cm}
\caption{(a-d) Thermal noise uncertainty (mK$^2$ Mpc$^3$). Angular modes are binned logarithmically, with 10 bins per decade, while line-of-sight modes are linear and set by the Fourier modes in the Fourier transform. Results from the same instrument (MWA or PAPER) are binned identically, have the same axis ranges, and the same minimum power level. (e) Expected 21~cm signal assuming a neutral medium at $z$=8 (see text for assumptions used in model). Parameters from Table \ref{instrument_design} are used for these results. Note the different color scales.}
\label{2d_thermal}
\end{figure}

\begin{figure}
\subfigure[MWA uniform $uv$ coverage.]{\includegraphics[scale=0.27]{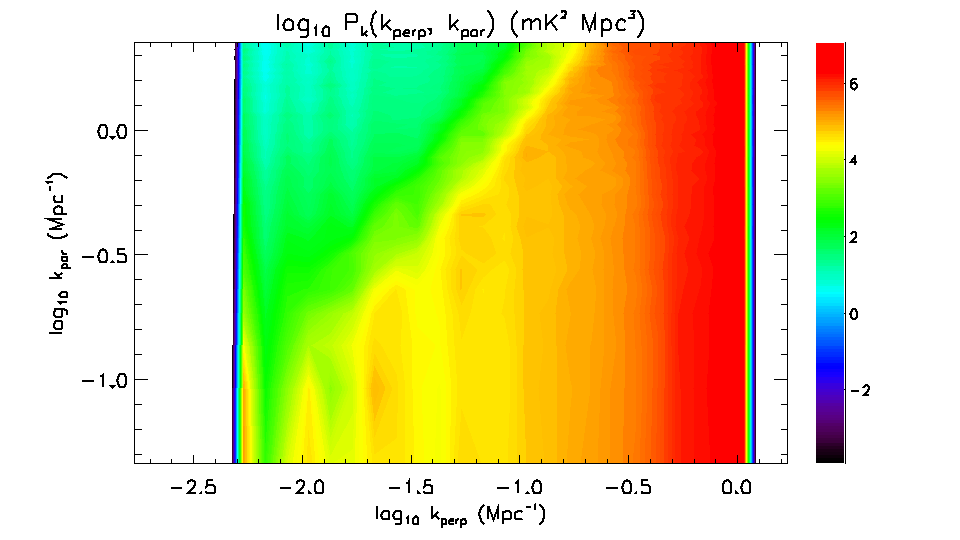}}
\subfigure[MWA.]{\includegraphics[scale=0.27]{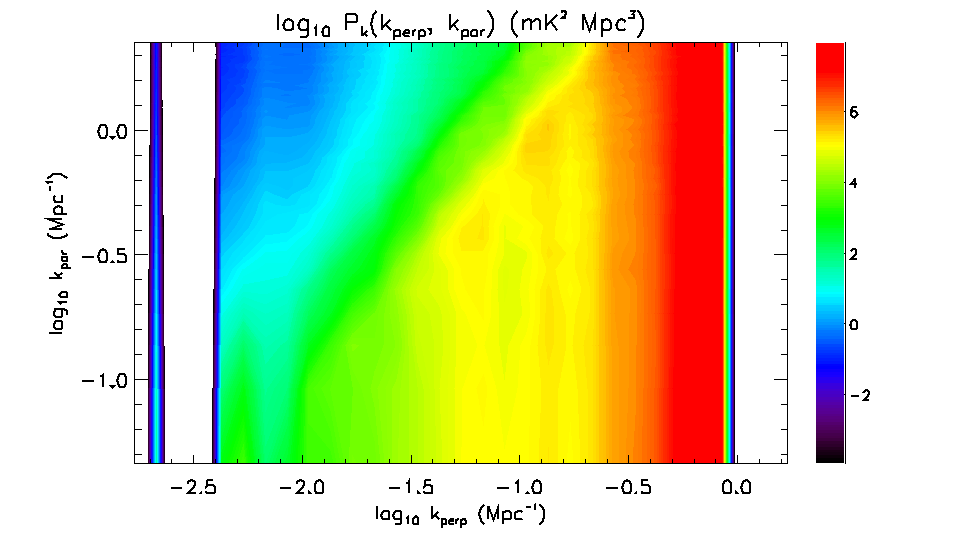}}\\
\subfigure[Actual 64-dipole minimum-redundancy (PAPER).]{\includegraphics[scale=0.27]{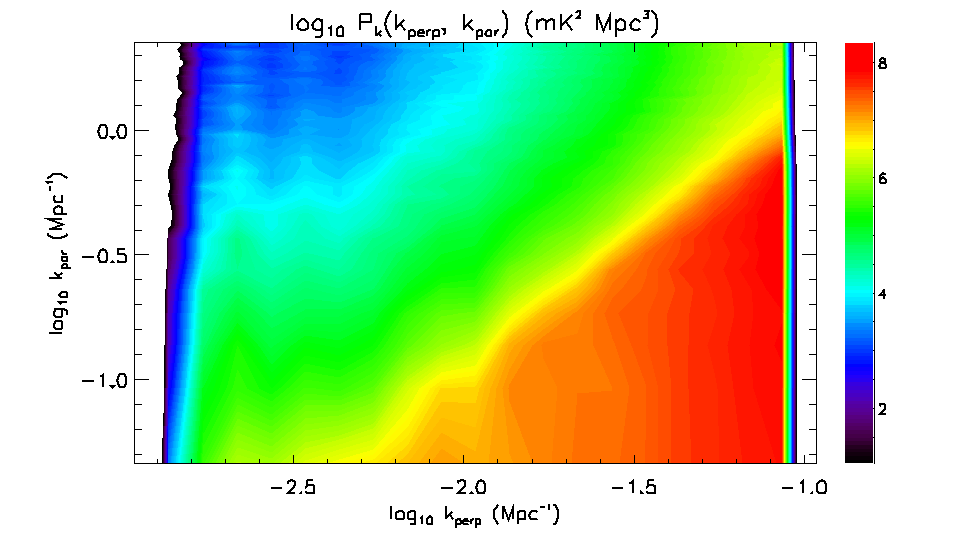}}
\subfigure[Potential 64-dipole maximum-redundancy (PAPER).]{\includegraphics[scale=0.27]{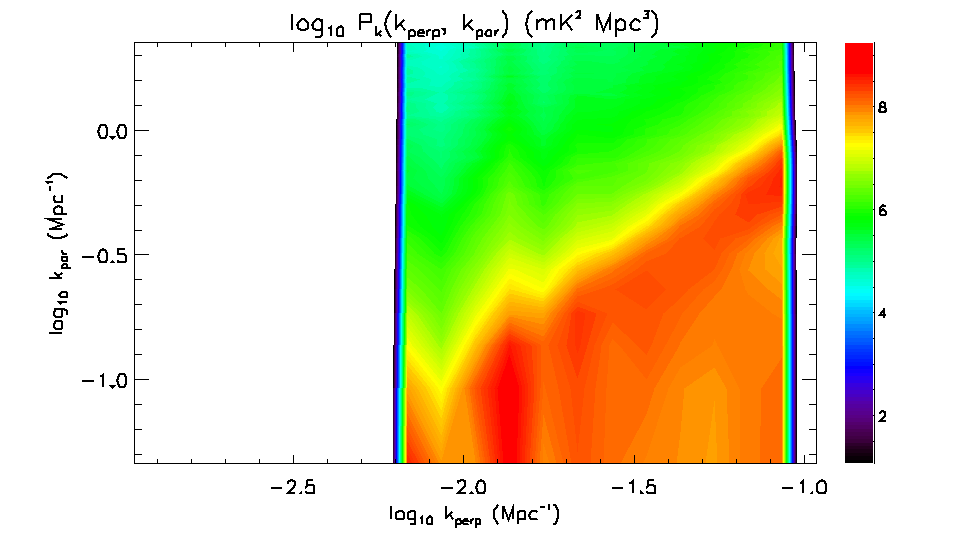}}
\vspace{1cm}
\caption{Residual signal uncertainty (mK$^2$ Mpc$^3$). Angular modes are binned logarithmically, with 10 bins per decade, while line-of-sight modes are linear and set by the Fourier modes in the Fourier transform. Results from the same instrument (MWA or PAPER) are binned identically, have the same axis ranges, and the same minimum power level. Parameters from Table \ref{instrument_design} are used for these results. Note the different color scales.}
\label{2d_residual}
\end{figure}

Figures \ref{2d_thermal}(a-d) and \ref{2d_residual} show the uncertainty in power for thermal noise and residual point source signal (mK$^2$ Mpc$^3$), respectively, for the four experiments considered, and obtained via Monte-Carlo simulation. Results from the same instrument (MWA or PAPER) are binned identically, have the same axis ranges, and the same minimum power level, but color scales differ. In addition to the thermal uncertainty, figure \ref{2d_thermal}(e) displays the expected signal from the 21~cm emission, assuming a fully-neutral IGM at $z$=8, where we have used a linear power spectrum, and the transfer functions of \citet{eisenstein99}, and assumed isotropy \citep{furlanetto06,mcquinn06}.

The thermal noise (Figure \ref{2d_thermal}) is uncorrelated between frequency channels, yielding the same expected power and uncertainty along the line-of-sight ($k_\parallel$), where the number of contributing visibilities is equal. On angular modes, the uncertainty is set by the number of visibilities within a $(uv)$ cell, and the binning used in the plotting (i.e., the coherent versus incoherent balance, and the size of the bins).  The same behaviour as seen for the angular power spectrum is observed: minimally-redundant arrays (Figure \ref{2d_thermal}a,c) yield flat power over a wide range of angular modes, while the centrally-concentrated/maximally-reduandant arrays (Figure \ref{2d_thermal}b,d) sample some modes particularly well, but at the expense of overall performance and range of sampled modes.

The residual signal uncertainty (Figure \ref{2d_residual}) displays qualitatively different behaviour to the thermal noise, owing to the correlated frequency channels and the dependence of noise variance on $|\boldsymbol{u}|$. The wedge-shaped feature extending from small $k_\bot$--small $k_\parallel$ to large $k_\bot$--large $k_\parallel$ was first reported by \citet{datta10}, after initial indications from \citet{bowman09}, and has been further discussed by \citet{vedantham12,morales12}, and is the power footprint of the instrument PSF. In Section \ref{wedge_section} and Appendix \ref{wedge_appendix} we explore the origin of this feature in more detail, and provide a pedagogical derivation of its shape and location.

The residual signal power shows a clear ramping of signal towards smaller angular scales (large $k_\bot$), again resulting from the $|\boldsymbol{u}|^2$ dependence of visibility-based noise variance.

\begin{figure}
\subfigure[Thermal noise uncertainty, MWA.]{\includegraphics[scale=0.27]{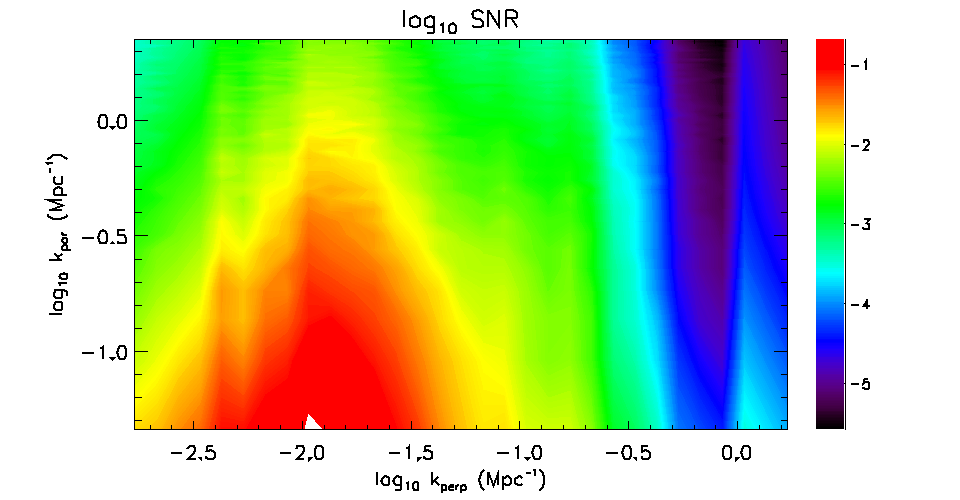}}
\subfigure[Residual signal uncertainty, MWA.]{\includegraphics[scale=0.27]{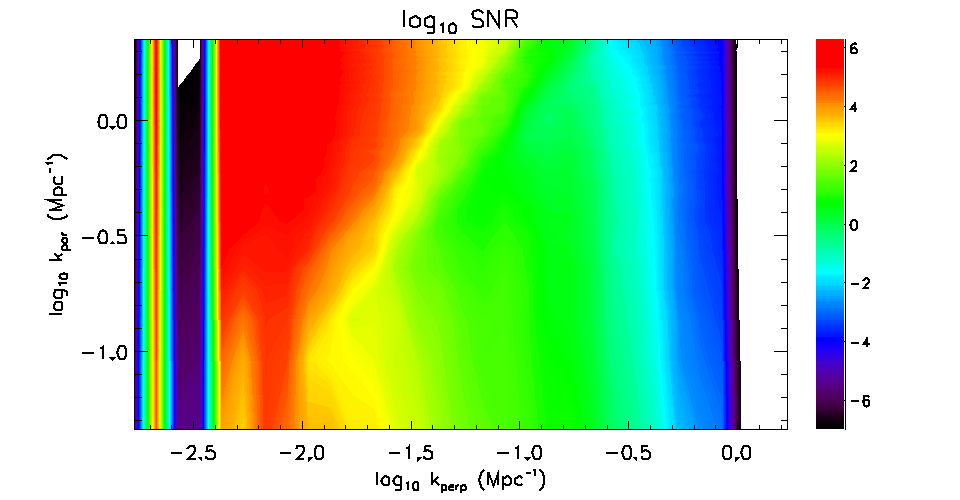}}\\
\caption{Signal-to-noise ratio of EoR statistical estimation assuming a fully-neutral IGM and isotropy, for (a) expected 21~cm signal to MWA thermal noise uncertainty, and (b) expected 21~cm signal to MWA residual signal uncertainty.}
\label{signal_to_noise}
\end{figure}

Figure \ref{signal_to_noise} displays the 21~cm signal-to-noise ratio (expected signal divided by noise uncertainty) for the actual MWA configuration, and for the thermal noise and residual signal contribution.
Consistent with previous estimates \citep{datta10,beardsley12}, we do not expect to detect the 21~cm EoR signal for 300 hours of observing and the parameters used here (SNR$<$1), when considering thermal noise. There is, however, a region of parameter space for which the signal-to-noise ratio exceeds 0.1, suggesting a possible detection in this region for $\sim$1000 hours (assuming coherent addition of visibilities). For the case of the residual point source signal, there is a large portion of parameter space, corresponding to high expected 21~cm signal and/or low uncertainty, where the SNR exceeds unity. These results are consistent with discussion isolating the low contamination region of $k_\bot-{k}_\parallel$ space for EoR detection and estimation.

\section{Explaining the ``wedge" feature}\label{wedge_section}
In Appendix \ref{wedge_appendix} we provide a full pedagogical motivation for the shape of the ``wedge" feature in the 2D power spectrum. Here we present the main results, and leave the interested reader to follow the full derivation in the Appendix.

The wedge feature is the footprint of the instrument PSF, transformed into power \citep{datta10,bowman09, vedantham12, morales12}. More specifically, it is a combination of two effects, which are then propagated through a Fourier transform, and squared to form power. The two effects are: (1) the integrated effect on frequency-based correlations of the subtraction of multiple point sources distributed randomly across the sky; and (2) the spread of frequency channels across $(uv)$ cells for visibilities observed with the same baseline (mode-mixing).

Residual signal from subtraction of a single point source from the visibilities yields frequency channels that are fully correlated in amplitude, but with a phase shift that is dependent on the source location and the baseline vector. Integrating the effect of multiple sources yields a sinc-like function in the frequency correlations, effectively reducing the frequency-based correlation length. In addition, the spread over coherent $(uv)$ cells of the visibilities forming a single baseline (mode-mixing) effectively truncates the sinc-like frequency correlations at the first sinc null. This effect is most pronounced on the longest baselines, and non-existent on the shortest (typically), where the visibilities from a given baseline all contribute to the same $(uv)$ cells. The frequency-based covariance matrix is therefore represented by a real-valued sinc function, multiplied by a rectangular window function, with width of the first sinc null. The Fourier transform and squaring of this functional form yields a wedge shape given approximately by:
\begin{equation}
W(k_\bot,k_\parallel) \propto {\rm sinc}^2\left(\frac{\alpha}{r_{\rm max}}\frac{k_\parallel}{k_\bot}\frac{\nu}{{\rm BW}} \right),
\label{sinc-squared1}
\end{equation}
where,
\begin{equation}
\alpha = \frac{\pi{c}(1+z)^2D_M(z)}{2H_0f_{21}E(z)},
\end{equation}
are array-independent terms defining the cosmology and observation redshift (with distance units), $r_{\rm max}$ is the instrument field-of-view (radial), BW is the experiment bandwidth, and $\nu$ is the centre frequency of the observation. Here $W$ represents the uncertainty in noise power (the expected power follows the same functional form) due to imprecise subtraction of multiple point sources, spread randomly across the instrument field-of-view.

We have therefore demonstrated that the wedge shape in the line-of-sight direction can be explained by a combination of mode-mixing (truncating the correlation length in frequency) and a real-valued sinc-like correlation function (derived from addition of random phases from individually-correlated point sources at random sky positions). In the angular modes, longer baselines are more severely affected than shorter baselines, with an approximate $|u|^2$ dependence of visibility variance (and consequently, expected power and power uncertainty).

\section{Discussion and conclusions}
Our results indicate that (1) optimal point source subtraction to 1~Jy will not be a limiting factor in statistical EoR estimation for proposed experiments with MWA and PAPER, contrary to the conclusions of previous analyses \citep{datta10}, which relied on assumed position errors that were independent of source strength; (2) frequency-channel covariance combined with spread of visibilities across coherent $(uv)$ cells yields the ``wedge" feature observed in point source subtracted EoR simulations; (3) angular mode correlations are expected to be small (1--2 orders of magnitude below variance level). We have presented an end-to-end derivation of the approximate shape of the wedge feature, using the first principles framework we have developed.
Here we describe some interesting features and caveats of these results, based on the assumptions within our framework.

The magnitude of the residual point source signal is dependent on the number of calibrators being peeled, the thermal noise level using the full array and bandwidth, and the angular mode in question (with a ${\sim}|\boldsymbol{u}|^2$-dependence on variance in the visibility). The uncertainty in the power spectrum at angular mode $l$ therefore scales with:
\begin{equation}
\sigma_{\rm PS}(l) \sim \frac{N_{\rm cal}l\sigma_{\rm therm}\Delta\nu}{\Delta\nu_{\rm tot}} \propto \frac{N_{\rm cal}l}{N_{\rm vis}\Delta{t}\Delta\nu_{\rm tot}},
\end{equation}
where $\sigma_{\rm therm}$ is the thermal noise level across the entire bandwidth, $\Delta\nu_{\rm tot}$, for power spectrum channels of width $\Delta\nu$. The magnitude of the expected power, and power uncertainty, therefore scale linearly with the number of peeled sources, and inversely with the total system bandwidth and experiment time. The PAPER array suffers from 1/4 the visibilities of the MWA and a larger field-of-view, but this is balanced somewhat by the additional bandwidth and the smaller $l$ modes available to the (more compact) instrument.

We have assumed a sequential peeling scheme, whereby sources are independently and sequentially estimated, and peeled from the dataset, without downstream impact on the estimation of the parameters of other sources. In reality, an experiment can choose to either (1) estimate the parameters of all sources simultaneously, whereby the approximation of pure radiometric noise in the dataset (ignoring calibration errors) is appropriate, or (2) estimate and peel sources sequentially, whereby the covariant noise matrix at each iteration is a combination of uncorrelated radiometric noise and correlated residual signal noise. The choice between these estimation techniques depends on the number of sources being peeled relative to the available data (number of unknowns versus constraints; the CRB is singular for underconstrained problems), and we leave this discussion for future work. If we ignore correlations between visibilities, a simple back-of-the-envelope calculation demonstrates that, for the parameters used in this work, our assumption is reasonable (where we have effectively expanded the noise covariance matrix as a Taylor series, and kept the first linear term). For subtraction of a larger number of sources, or for fewer visibilities, this approximation breaks down, and the compound effect of imprecisely estimating source parameters within ever-noisier data will increase the residual signal power towards the amplitude of the thermal noise power.

A related assumption in our model is the ignorance of previous calibration solutions for estimation of the parameters of a source. As demonstrated in Figure \ref{source_precision_plot}, optimal source position estimates with the full-bandwidth PAPER array (and an 8 second integration) have precision comparable to the potential ionospheric differential refraction magnitude, for sources close to the 1~Jy peeling floor. In these cases, the utility of the current dataset may be augmented by use of prior information about the local behaviour of the ionosphere. For larger arrays, this is also an issue if one attempts to peel below $\sim$1~Jy. The optimal method for estimating source location (balance of prior and current information, and how these are combined) will depend on the peeling floor, the efficiency of the estimation method, and the specific array design.

Discussion of realized estimation performance highlights a further basic assumption of our framework; we have proceeded from a fundamental reliance on the existence of an efficient and unbiased estimator (i.e., an estimation method that achieves the CRB on parameter precision). This has allowed us to calculate \textit{lower limits} on the impact of point source subtraction, and using a first principles approach, but does not address the issue of the existence of such a method (a method may not exist that can achieve the bound). In addition to designing an efficient estimator, the CRB applies only to unbiased estimators, and introducing bias into the estimation will yield different structures in the final power spectrum than shown here. As a simple estimate of the impact of an unbiased, but inefficient estimator, we (arbitrarily) assume we have estimated the source parameters to within three times the CRB, and calculate the residual signal uncertainty in the angular power spectrum. This toy model yields an increase in uncertainty of a factor of $\sim$10, indicating a rough square scaling of estimator inefficiency with noise uncertainty.

We have derived a ${\sim}|\boldsymbol{u}|^2$-dependence of visibility-based variance on baseline length, demonstrating a ramping of noise power at large angular modes (small angular scales). In this work, we have used the same antennas to estimate the source parameters and produce the power spectra (the only difference being the bandwidths used). The long baselines contribute relatively more than the short baselines to the estimate of source position (equations \ref{l_precision}--\ref{m_precision}). However, these baselines form the largest $l$ modes, where residual signal has the greatest impact. Inclusion of these baselines for parameter estimation is crucial for precise estimates; their use in EoR measurements depends on the experimental design, and the desire for measuring the 21~cm EoR signal at those scales. Use of these baselines does not affect EoR estimation at other scales (the correlation between modes is small). Also, as demonstrated in \citet{trott11}, inclusion of short baselines for position estimation is important (but less so than long baselines), due primarily to the increased instrumental sensitivity when using all available antennas.

Our results also demonstrate the expected behaviour of minimally-redundant versus maximally-redundant arrays. Uniform sampling yields even thermal noise power across a wide range of angular modes, while maximally-redundant arrays sample some scales well, but yield patchy performance and restricted range of sampled modes. Choice between these models depends on the experiment being performed, but a maximally-redundant array requires prior information about the most interesting scales for estimating the 21~cm signal, while the uniform arrays are a better ``starter" design, where the scientist has no prior information about signal structure and is aiming for a first detection/estimation.

\section*{Appendix}
\appendix
\section{Scaling of expected power spectrum and variance with number of visibilities}
Computation of the angular power spectrum involves both a coherent and incoherent addition of visibilities. The relative size of the coherent to incoherent calculation depends on the number of visibilities contributing to each ($uv$)-cell, and the number of ($uv$) cells contributing to each $l$-mode. In general, the relative size of these additions is different for each $l$-mode. As such, the scaling of the expected value of the point source residual power spectrum, and its variance, with the number of contributing visibilities is non-trivial, and the design of the interferometer, and the binning of the power spectrum, affect the result. In this Appendix, we will briefly outline how the relative size of the coherent to incoherent combination scales with the number of contributing visibilities, beginning with the two extreme examples (all coherent and all incoherent), and then for an intermediate example. This is motivated by the discussion of minimum-redundancy versus maximum-redundancy configurations of \citet{parsons11}, and the initial use of the PAPER array in both types of configurations. Typically, visibilities can be considered coherent over scales that approximate the instrument field-of-view. Arrays that sample the same angular modes coherently (maximally-redundant) can reduce the contribution of thermal noise on these scales, at the expense of being able to sample a wider range of angular scales (for a fixed number of antennas).

The angular power spectrum is estimated using a maximum-likelihood estimator, designed to weight the contributions from each cell according to the noise they contain. The estimate for mode $l$ is given by;
\begin{equation}
C_l = \frac{\displaystyle\sum_{(uv) \in l}N_{uv}|V_{uv}|^2}{\displaystyle\sum_{(uv) \in l}N_{uv}},
\end{equation}
where $N_{uv}$ is the number of visibilities contributing to a given ($uv$) cell, and the sum is over the ($uv$)-cells contributing to that $l$-mode. The visibilities, $V_{uv}$, are \textit{coherently} combined (averaged) within each ($uv$) cell, while the final estimate associating ($uv$)-cells in $l$-modes is \textit{incoherent} (visibilities are combined after squaring). For simplicity, we will compute the statistics for complex white Gaussian noise on each visibility, distributed as $V \sim \mathcal{CN}(0,\sigma^2)$ (i.e., the thermal noise, where `$\mathcal{CN}\sim(0,\sigma^2)$' denotes that the PDF is distributed according to a zero mean complex Gaussian distribution with variance, $\sigma^2$).

For an entirely coherent estimate ($N_l = N_{uv}$), where a single ($uv$) cell contributes to a single $l$-mode, the summations are omitted;
\begin{equation}
C_l = |V_{uv}|^2,
\end{equation}
yielding,
\begin{equation}
C_l \sim \mathcal{N}\left(\frac{\sigma^2}{N_l},\frac{2\sigma^4}{N_l^2} \right).
\end{equation}
Note that in this case, increasing the number of visibilities reduces the expected noise power and variance, but the ratio of expected noise power to standard deviation of power is unchanged. Strictly speaking, this is a chi-square distribution with one degree of freedom, yielding a highly skewed probability distribution function, $f$,
\begin{equation}
f(C_l) = \frac{\exp{(-\frac{x}{2})}}{\sqrt{2x}\Gamma{(\onehalf)}},
\end{equation}
where $x = C_l\sigma^2/N_l$.

At the opposite extreme, for an entirely incoherent estimate ($N_{uv}=1$), where there is a single visibility contributing to each ($uv$) cell, the power spectrum estimate is given by:
\begin{equation}
C_l = \frac{\displaystyle\sum_{i=1}^{N_l}|V_i|^2}{N_{l}},
\end{equation}
yielding,
\begin{equation}
C_l \sim \mathcal{N}\left({\sigma^2},\frac{2\sigma^4}{N_l} \right).
\end{equation}

In the intermediate case, where we would expect to be for real experiments, there will be large numbers of visibilities combined coherently, and a mapping of 10s-100s ($uv$) cells to each $l$ mode. We take an example with ten ($uv$) cells contributing to a given $l$-mode, with an even number of visibilities in each ($uv$)-cell, $N_l = 10N_{uv}$. The power spectrum estimate is given by:
\begin{equation}
C_l = \frac{\displaystyle\sum_{i=1}^{10} N_{uv}|V_{uv}|^2}{\displaystyle\sum_{i=1}^{10} N_{uv}},
\end{equation}
yielding,
\begin{equation}
C_l \sim \mathcal{N}\left(\frac{\sigma^2}{N_{uv}},\frac{2\sigma^4}{N_lN_{uv}} \right).
\end{equation}
This expression forms the natural transition from completely coherent to completely incoherent, and we have again taken the Gaussian approximation to the chi-square distribution.

\section{Pedagogical motivation for the `wedge' feature}
\label{wedge_appendix}
The `wedge' of residual power expected from point source subtraction has been discussed by \citet{datta10}, \citet{vedantham12}, and \citet{morales12}, and represents the power footprint of the instrument point spread function. In particular, the review of \citet{morales12} presents arguments for the origin of this feature due to point source amplitude and position errors, gridding artifacts and calibration errors. They motivate the position and structure of the wedge, but without a first-principles approach to the magnitude of the errors. In this Appendix, we extend the previous discussion by (1) motivating the shape of the wedge from the first-principles approach followed here, providing an alternative perspective on its shape and location, and (2) complementing previous descriptions by describing the \textit{integrated} effect of multiple imprecisely-subtracted point sources. Note that we are not suggesting previous analyses are incorrect; we are simply approaching the problem from a different path.

We have already demonstrated that with an unbiased and efficient estimate of the parameters of each point source (i.e., obtaining the CRB on parameter precision), the error introduced into a given visibility scales as the square of the baseline length (see Equations \ref{error_vis3} in Section \ref{error_prop_section}). This analysis explains the ramping of the residual signal power in the $k_\bot$ dimension. In this Appendix we focus on the shape in the $k_\parallel$ direction; specifically, the sharp drop-off of signal from small to large $k_\parallel$ at large angular scales (small $k_\bot$), and the relatively flat signal power at small angular scales (large $k_\bot$). We will demonstrate that this structure is a combination of correlated signal across frequency channels spreading across multiple $uv$-cells (``mode-mixing"), and the integrated effect of subtracting multiple sources distributed randomly across the sky.

For a single source contaminating a $(u,v,\nu)$ cell, the magnitude of the covariance between frequency channels (for the same $(u,v)$) is equal to unity (i.e., they are completely correlated). There is, however, a phase shift in the complex covariance matrix, yielding the following structure (equation \ref{jacobian}):
\small
\begin{eqnarray}
[\boldsymbol{C}_{uv}(\nu)] &=& {\sigma_V^2}
\begin{pmatrix}
1 & \exp{i\phi} & \exp{2i\phi} & \cdots & \exp{Ni\phi} \\
\exp{-i\phi} & 1 & \exp{i\phi} & \cdots & \exp{(N-1)i\phi} \\
\exp{-2i\phi} & \exp{-i\phi} & 1 & \cdots & \exp{(N-2)i\phi} \\
\vdots & \vdots & \vdots & \ddots & \vdots \\
\exp{-Ni\phi} & \exp{-(N-1)i\phi} & \exp{-(N-2)i\phi} & \cdots & 1
\label{matrix}
\end{pmatrix}
\end{eqnarray}
\normalsize
where
\begin{equation}
\phi = 2\pi\Delta\nu(\Delta{x}l+\Delta{y}m)/c
\label{phimax}
\end{equation}
is the phase term over frequency range $\Delta\nu$ (a single channel) for a source located at sky position $(l,m)$ and baseline lengths $(\Delta{x},\Delta{y})$, and $\sigma_V^2$ is the source-independent variance in a visibility, given by equation \ref{error_vis3}. This equation can be derived using the off-diagonal entries of equation \ref{jacobian}.

When there are multiple sources contributing residual signal to the coherently-averaged visibility, these phase terms add, yielding a sum over independent phases for the matrix entry describing the covariance between frequency channels, $\nu_\alpha,\nu_\beta$:
\begin{equation}
\boldsymbol{C}_{uv}(\nu_\alpha,\nu_\beta) = \displaystyle\sum_{j=1}^{N_{\rm cal}} \exp{(\alpha-\beta)i\phi_j}.
\end{equation}
For phases distributed evenly over $2\pi$, the expected value of this sum is zero, and the frequency channels would be completely uncorrelated. However, the visibility phase term, $\phi$, is limited in magnitude by the extent of the array (maximum ($\Delta{x},\Delta{y}$)) and the field-of-view (maximum $(l,m)$). If we assume, for simplification, that the phases are uniformly-distributed over some range, $[-\theta,\theta]$ (corresponding to randomly-located sources and a uniform density of antennas), the expected value of the summed phases is:
\begin{equation}
E\left[\displaystyle\sum_{j=1}^{N_{\rm cal}} \exp{i\phi_j} \right] = N_{\rm cal}{\rm sinc}\hspace{1mm}{\theta},
\end{equation}
where $\theta=\phi_{\rm max}$, and the variances (matrix diagonal entries) have also summed to $N_{\rm cal}$. Note that the expected value of the off-diagonal terms sum to a real quantity. Hence, the covariance matrix, summed over multiple sources, is real, with a correlation between frequency channels that decays as a sinc function with an argument that is linear in the frequency spacing between the channels under consideration. The maximum value of $\phi$ is determined by the instrument field-of-view and the baseline length under consideration. At small values of $|\boldsymbol{u}|^2$ (i.e., at small $k_\bot$ values) the maximum value is smaller than at larger baseline lengths (large $k_\bot$ values). Hence, the correlation between frequency channels decays more rapidly at small angular scales (large $k_\bot$). For the MWA, with a maximum baseline of 3000~m (minimum 5~m), a field-of-view of 30$^o$, and frequency channels of 80~kHz (used in this work), the maximum values of $\phi$ for the shortest and longest baseline are $\sim$0.002 and $\sim$1.3, respectively. Therefore, on the longest baselines, the sinc function reaches its first null after a few channels, while on the shortest baselines, it is $>$1000 (which exceeds the number of channels considered for a 8~MHz bandwidth).

In addition to the integrated effect of imprecisely subtracting point sources, the noise structure of the power spectrum is affected by mode-mixing: the evolution of the instrument PSF with frequency, which manifests itself in distributing the signal from a given baseline over several $(uv)$ cells. In the context of covariances between frequency channels in the same $(uv)$ cell, this straying of visibilities leads to a cutoff in the correlation length, dependent upon the width of the coherent $(uv)$ cell relative to gradient in $(uv)$ evolution with frequency. Mode-mixing is negligible on short baselines, and can be considerable on long baselines. The size of a coherent $(uv)$ cell is given approximately by the inverse of the instrument field-of-view, yielding a $\Delta{u}_{\rm cell}\sim{2}$ cell-size for the MWA with a 30$^o$ field. The spread of $u$ values for a given baseline is given by $\Delta{u} = \Delta{x}{\rm BW}/c$, which has the range [0.1,80] for the MWA. Hence, at the shortest baselines, all visibilities are contained within a single cell, while they begin to shift cells for $\Delta{x} \gtrsim \Delta{u}_{\rm cell}c/{\rm BW} \approx 70$~m (with the longest baselines traversing $>$40 cells).

Armed with an understanding of the structure of the frequency-based covariance matrix, we can now explore how the covariances propagate through the $\nu-\eta$ Fourier transform and power (mod-squared) operations, and lead to the wedge feature observed in the results. For an uncorrelated dataset, the variance in the power formed via DFT and squaring is the same for all Fourier modes ($\eta$), even for different variances in the individual frequency-based dataset, and is given by the sum of the squares of the variances of each component (equation \ref{quadratic_form_eqn}):
\begin{equation}
\sigma^2_{\eta} = 2{\rm tr}(\boldsymbol{AC^{\dagger}AC}) = \displaystyle\sum_{i=1}^{N_{ch}}\sigma^4_i,
\end{equation}
where $N_{ch}=N_\eta$ is the number of frequency channels and Fourier bins. This behaviour corresponds to the long baselines, where the combination of mode-mixing and small frequency correlation length yield relatively uncorrelated matrices. Visibilities on the longest baselines traverse cells after $\sim{2}-3$ frequency channels, yielding a substantially truncated channel correlation, and effectively uncorrelated frequency channels. At the other extreme, the larger angular modes (small $k_\bot$, $uv$) suffer from little mode-mixing and have longer correlation lengths, yielding qualitatively different behaviour in the DFT$+$squaring operations.

We now motivate the shape of the wedge as a function of $k_\bot$. We have demonstrated that the frequency channels are correlated with a sinc function dependence, yielding a visibility dataset that corresponds to Gaussian noise convolved with a sinc function. The Fourier transform of this convolution yields a flat line-of-sight power, multiplied by the Fourier transform of a sinc function (a rectangle function). I.e., it performs a low-pass filtering operation. Hence, the expected noise power follows a rectangle function (squared), with a width determined by the correlation length in the frequency domain:
\begin{eqnarray}
&&F[\mathcal{N}(0,\sigma^2) \star {\rm sinc}(\phi_{\rm max}n)]\\
&=& F[\mathcal{N}(0,\sigma^2)] . F[{\rm sinc}(\phi_{\rm max}n)]\\
&\propto& F[{\rm sinc}(\phi_{\rm max}n)]\\
&\propto& {\rm rect}\left(\frac{\pi{k}}{N\phi_{\rm max}}\right),
\end{eqnarray}
where
\begin{eqnarray}
{\rm rect}(x) = \begin{cases} 1 &\mbox{if } x \geq 1 \\
0 & \mbox{if } x < 1 \end{cases}
\end{eqnarray}
is the rectangle function, and $n,k = [0,1,..,N-1]$ index the DFT. This function yields a constant expected power when $k=N$, and a single bin rectangle function when $k=1$, producing the following limits on baseline lengths (using equation \ref{phimax}):
\begin{eqnarray}
\Delta{x} \lesssim 70{\rm m} &\mbox{if } k=1 \\
\Delta{x} \sim 7000{\rm m} & \mbox{if } k=N .
\end{eqnarray}
The latter baseline length exceeds the maximum for the MWA. One would therefore expect that the noise power would not be flat across the line-of-sight range, but instead fall off at large $k_\parallel$. This analysis neglects the mode-mixing, however, and the sinc function (and therefore the correlation length) is severely truncated for the longest baselines. The sinc function is always truncated at its first null (argument=$\pi$), because this corresponds to the scale over which the visibilities for a given baseline traverse the coherent $(uv)$ cell. This can be seen by considering the number of frequency channels to reach the first sinc function null, and the number of frequency channels over which a baseline is contained within a single cell. In one dimension:
\begin{eqnarray}
&N_{\Delta\nu,{\rm sinc}} = \frac{\pi}{\phi_{\rm max}} = \frac{c}{2\Delta\nu{r}_{\rm max}\Delta{x}}\\
&N_{\Delta\nu,{\rm mm}} = \frac{\Delta{u}_{\rm cell}}{\Delta{u}_{\Delta\nu}} = \frac{1}{2r_{\rm max}}\frac{c}{\Delta\nu\Delta{x}},
\end{eqnarray}
where `mm' refers to mode-mixing. These two expressions are seen to be equivalent. The truncation windows the frequency-domain sinc correlation with a rectangle function, tapering the power-domain rectangle function with a sinc-like function:
\begin{eqnarray}
&& \left | \mathcal{F}\left[ {\rm sinc}(n\phi)\right] . \left[{\rm rect}\left(\frac{n\phi}{\pi} \right) \right] \right |^2 \\
&\propto& \left | {\rm rect}\left(\frac{\pi{k}}{N_{ch}\phi} \right) \star {\rm sinc}\left(\frac{\pi^2{k}}{N_{ch}\phi} \right) \right |^2 \\
&\simeq& {\rm sinc}^2\left(\frac{\pi^2{k}}{N_{ch}\phi} \right),
\end{eqnarray}
where $k$ denotes the line-of-sight Fourier bin. Inserting the expression for $\phi=\phi_{max}$ and transforming co-ordinates to $k_\bot$ and $k_\parallel$ yields the following approximate expression for the wedge structure:
\begin{equation}
W(k_\bot,k_\parallel) \propto {\rm sinc}^2\left(\frac{\alpha}{r_{\rm max}}\frac{k_\parallel}{k_\bot}\frac{\nu}{{\rm BW}} \right),
\label{sinc-squared}
\end{equation}
where,
\begin{equation}
\alpha = \frac{\pi{c}(1+z)^2D_M(z)}{2H_0f_{21}E(z)},
\end{equation}
are array-independent terms defining the cosmology and observation redshift (with distance units), BW is the experiment bandwidth at observation frequency, $\nu$. Here $W$ represents the uncertainty in noise power (the expected power follows the same functional form) due to imprecise subtraction of multiple point sources, spread randomly across the instrument field-of-view.

\begin{figure}[ht]
{\includegraphics[scale=0.9]{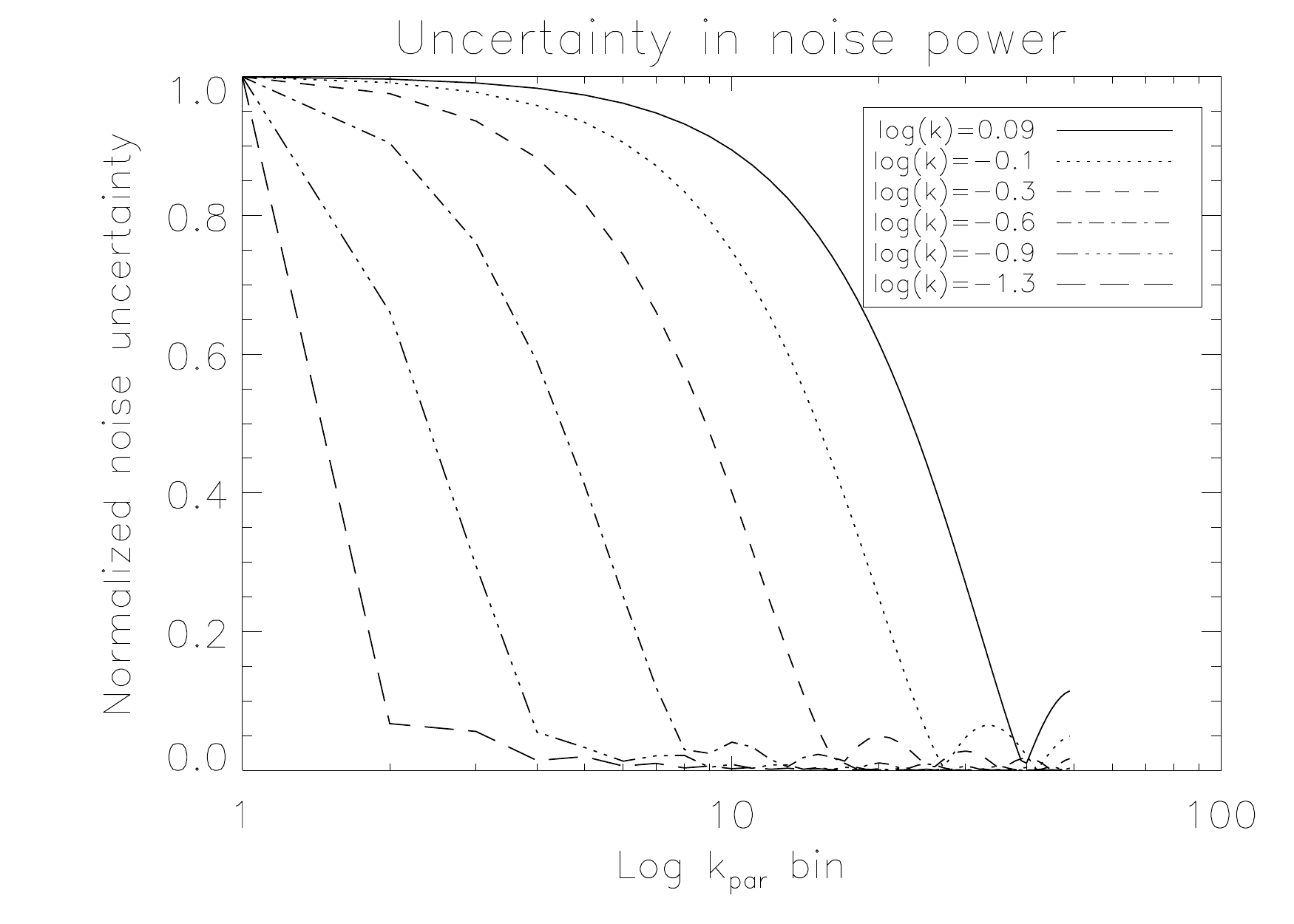}}
\vspace{1cm}
\caption{Uncertainty in noise power for different values of $k_\bot$ for the MWA (normalized to first non-DC term in the power).}
\label{sinc_func}
\end{figure}

We have therefore arrived at the final form of the frequency covariance matrix --- a real-valued truncated sinc function, with a truncation length at the first null --- yielding a sinc-squared-like function with characteristic length given by the argument of equation \ref{sinc-squared}. Figure \ref{sinc_func} plots examples of this function, as a function of line-of-sight Fourier bin, for different values of $k_\bot$.
The analytic result mimics the qualitative structure observed in the power spectra shown in this work, and in previous published results (the result is approximate because it relies on simplifying assumptions, namely that the correlation length is truncated exactly at the first null of the sinc function, and that the antenna layout is such to produce a uniform sampling of correlation phases in the frequency covariance matrix).

\acknowledgments{The authors would like to thank Aaron Parsons, Jonathan Pober, Danny Jacobs, and the PAPER collaboration for providing the PAPER antenna configurations, and Steven Tingay and Stephen Ord for useful discussions. The Centre for All-sky Astrophysics is an Australian Research Council Centre of Excellence, funded by grant CE11E0090. The International Centre for Radio Astronomy Research (ICRAR) is a Joint Venture between Curtin University and the University of Western Australia, funded by the State Government of Western Australia and the Joint Venture partners. RBW is supported via the Western Australian Centre of Excellence in Radio Astronomy Science and Engineering.}

\newpage
\bibliographystyle{jphysicsB}
\bibliography{paper.bib}

\end{document}